\documentclass[twocolumn]{aastex6}

\fullcollaborationName{The Friends of AASTeX Collaboration}

\def\solm{M$_{\odot}\,$}

\def\solm{M$_{\odot}\,$}

\def\casgm20{CAS-G-M$_{20}\,$}
\def\m20{M$_{20}\,$}

\begin{document}


\title{The Evolution of Galaxy Number Density at $z < 8$ and its Implications}



\author{Christopher J. Conselice, Aaron Wilkinson, Kenneth Duncan\altaffilmark{1}, Alice Mortlock\altaffilmark{2}}
\affil{University of Nottingham, School of Physics \& Astronomy, Nottingham, NG7 2RD UK}

\altaffiltext{1}{Leiden Observatory, Leiden University, PO Box 9513, 2300 RA Leiden, the Netherlands}
\altaffiltext{2}{SUPA, Institute for Astronomy, University of Edinburgh, Royal Observatory, Edinburgh, EH9 3HJ}

\and


\begin{abstract}

The evolution of the number density of galaxies in the universe, and thus also the total number of galaxies, is a fundamental question with implications for a host of astrophysical problems including galaxy evolution and cosmology. However there has never been a detailed study of this important measurement, nor a clear path to answer it. To address this we use observed galaxy stellar mass functions up to $z\sim8$ to determine how the number densities of galaxies changes as a function of time and mass limit. We show that the increase in the total number density of galaxies ($\phi_{\rm T}$), more massive than M$_{*} = 10^{6}$ \solm, decreases as $\phi_{\rm T} \sim t^{-1}$, where $t$ is the age of the universe. We further show that this evolution turns-over and rather increases with time at higher mass lower limits of M$_{*}>10^{7}$ \solm. By using the M$_{*}=10^{6}$ \solm lower limit we further show that the total number of galaxies in the universe up to $z = 8$ is $2.0^{+0.7}_{-0.6} \times 10^{12}$ (two trillion), almost a factor of ten higher than would be seen in an all sky survey
at Hubble Ultra-Deep Field depth. We discuss the implications for these results for galaxy evolution, as well as compare our results with the latest models of galaxy formation. These results also reveal that the cosmic background light in the optical and near-infrared likely arise from these unobserved faint galaxies. We also show how these results solve the question of why the sky at night is dark, otherwise known as Olbers' paradox.

\end{abstract}


\keywords{Galaxies:  Evolution, Formation, Structure, Morphology, Classification}



\section{Introduction}

When discovering the universe and its properties we are always 
interested in knowing absolutes.  For example, it is of astronomical
interest to calculate how many stars are in our Galaxy, how many 
planets are surrounding these stars (Fressin et al. 2013), the total 
mass density of the universe (e.g., Fukugita \& Peebles 2004), amongst  
other  absolutes in the universe's properties.  
One  of these that has only been answered in a rough way is 
the total number density evolution of galaxies, and thus
also the total number of galaxies in the universe.

This question is not only of passing interest as a curiosity, but is also 
connected to many other questions in cosmology and astronomy.  The evolution
of the number 
densities of galaxies relates to issues such as galaxy formation/evolution 
through the number of systems formed, the evolution of the
 ratio of giant galaxies to dwarf galaxies, the distant supernova and 
gamma-ray burst rate, the star formation rate of the universe,
 and how new galaxies are created/destroyed through 
mergers (e.g., Bridge et al. 2007; 
Lin et al. 2008; Jogee et al. 2009; Conselice et al. 2011; 
Bluck et al. 2012;
Conselice 2014; Ownsworth et al. 2014).  The number of galaxies in the
observable universe also divulges information about the mass density of the universe,  background light at various wavelengths, as well as insights 
into Olbers' Paradox.  However, there still does not yet exist a good
measurement for this fundamental quantity.

Understanding the co-moving number density evolution of galaxies has only been 
possible in any meaningful way
since deep imaging with telescopes began with the advent of CCD cameras.  Deep
surveys to search for distant galaxies started in the 1990s (e.g., Koo \& Kron
1992; Steidel \& Hamilton 1992; Djorgovski et al. 1995), and reached our 
current depths after deep Hubble Space Telescope imaging campaigns were
carried out, especially within the Hubble Deep Field 
(Williams et al. 1996).  This was later expanded to other deep fields such
as the Hubble Deep Field South (Williams et al. 2000), the Great Observatories
Origins Survey (Giavalisco et al. 2004), and the near-infrared
CANDELS fields (Grogin et al. 2011; Koekemoer et al. 2011), and finally to the Hubble Ultra Deep
Field (Beckwith et al. 2006) which remains the deepest image in the optical
and near-infrared of our universe taken to date.  

However, despite these surveys it is still uncertain how the total number
density of galaxies evolves over time.  This is an interesting question
as we know that the star formation rate rises, and then declines at $z < 8$
(e.g., Bouwens et al. 2009; Duncan et al. 2014; Madau \& Dickinson 2014), 
while at the same time
galaxies become larger  and less peculiar (e.g., Conselice
et al. 2004; Papovich et al. 2005; Buitrago et al. 2013; 
Mortlock et al. 2013; Lee et al. 2013; Conselice 2014; Boada et al. 2015).
However, we do not know how the total number of galaxies at a given 
epoch evolves, and how this is associated with 
the general formation of the galaxy population as a whole.

There are a few reasons for why deep imaging programs are not easily
able to convert observations to total numbers of galaxies. One of these 
issues is that all deep observations are incomplete. This is due to 
limitations in exposure times and depth such that certain galaxies will be 
detected more readily than other galaxies.
The result of this is an incompleteness down to the magnitude limit of 
even the deepest
surveys, which can be corrected for, but which still leaves
some uncertainty.   However, the more important issue is that these 
observations do not reach the faintest galaxies, although from number
density fits and theory we know that there should be many more faint galaxies
beyond our current observational limits.

It is also important to address what we mean by the total number density of 
galaxies in the universe.   This is not a simple quantity to define as
the total number density which exists now, the total number density 
which is observable
in principle, and the total number density which is observable with current
technology, are all different questions with different answers. 
There is also the issue that we are limited by the cosmological 
horizon over what we can observe, and therefore there are galaxies
we cannot see beyond it. Even the number of galaxies which exist in the
universe today, i.e., if we could view the entire universe as is, and not
be limited by light travel time,  is a complicated question. Galaxies in  
the distant universe have evolved beyond what we can currently observe due 
to the finite nature of the speed of light, and presumably would look similar
to those in the local universe.  We address these issues in 
the paper. Our default and ultimate total number density of galaxies
we investigate in this paper is how the
number density evolves within the current observable universe up to $z \sim 8$.

For comparison purposes, we also carry out an analysis in the Appendix of
the number of galaxies which are visible to modern telescopes, at all 
wavelengths, that we can currently observe.  We then compare this to 
measurements of the total number that actually can be potentially observed 
in the universe based on measured mass functions.  
We also discuss how these results reveal information concerning 
galaxy evolution and background light.  We
also give indications for future surveys and what fraction of galaxies
these will observe.

 
This paper is divided up into several sections.  \S 2 describes the data 
we use throughout this analysis, \S 3 describes the results of this paper
including using  fits of galaxy stellar mass functions to 
derive the total number of galaxies which are
in the universe.  \S 4 describes the implications of these results and
\S 5 is a summary. Throughout this paper we use a standard 
cosmology of H$_{0} = 70$ km s$^{-1}$ Mpc$^{-1}$, and 
$\Omega_{\rm m} = 1 - \Omega_{\lambda}$ = 0.3.

\section{Data}

The data we use for this paper come from many sources, and 
results from previous papers.  In the Appendix we
describe how many galaxies we can actually observe at present in the 
universe, based on the deepest observations yet available.  Here 
in the main paper we 
address the question of how many galaxies are potentially detectable within 
the universe if deep imaging, over all wavelengths, could be carried out 
in every location of the sky without any interference from the Galaxy, 
or other contamination.

\begin{figure*}
 \vbox to 85mm{
\includegraphics[angle=0, width=180mm]{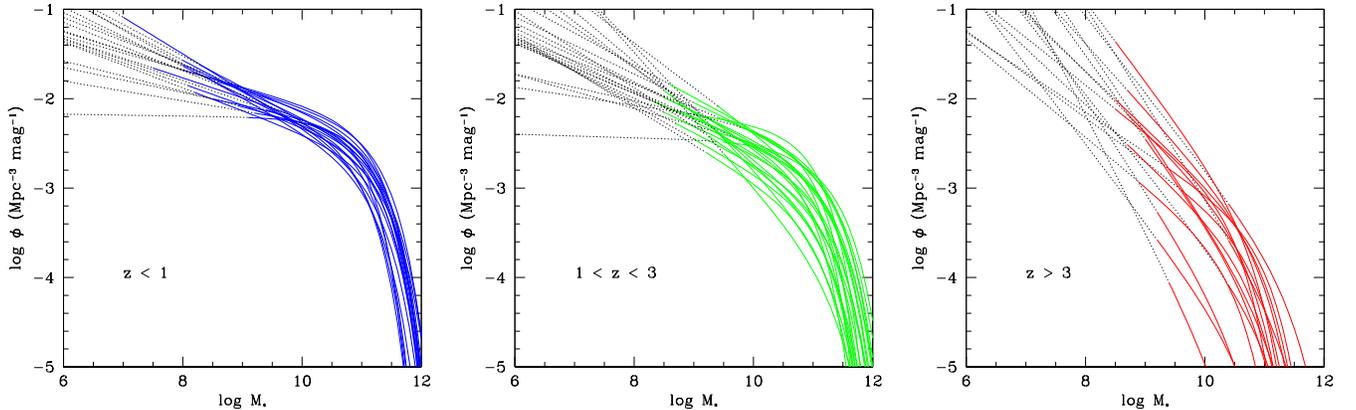}
 \caption{The mass functions which we use throughout this paper, plotted
through  best fitting Schechter function fits.  These
are all from the various studies described in \S 2.  The mass functions are
shown in terms of redshifts, such that the left panel shows systems at 
$z < 1$, the
middle panel shows $1 < z < 3$ and $z > 3$ (far right).  These mass functions
are shown such that the solid colored lines are the mass functions down to the limit
of the respective data whereby it is complete, and the dotted lines show our
extrapolation down to M$_{*} = 10^{6}$ \solm.  The `flattest' mass function
plot in the $1 < z < 3$ regime is from Muzzin et al. (2013) and the flattest
$z > 3$ mass function is from Grazian et al. (2015). }
\vspace{-6cm}
} \label{sample-figure}
\end{figure*}

For the bulk analysis and results of this paper we utilize mass functions 
of galaxies from the local universe up until $z \sim 8$ to determine how 
the number
density of galaxies evolves with time and redshift.     These mass 
and luminosity functions are now just starting to be measured at 
high redshfits, and our primary data originates from mass functions calculated 
using deep and wide near-infrared  and optical surveys with Hubble and ground
based telescopes.  

As described in the next section, the mass functions we utilize are from 
Perez-Gonzalez et al. (2008), Kajisawa et al. (2009), Fontanta et al.
(2004, 2006), Caputi et al. (2011), Pozzetti et al. (2007),
Mortlock et al. (2011), Tomczak et al. (2014), Muzzin et al. (2013), 
and Mortlock et al. (2015) for galaxies at $z < 3$.  At the highest redshifts
we use mass functions published in Duncan et al. (2014), Grazian et al. 
(2015), Caputi et al. (2011) and Song et al. (2015).    We normalize
all of  these mass
functions for each survey based on the Salpeter initial mass function (IMF)
for stars between 0.1 \solm and 100 \solm.   We also use only the 
co-moving number
densities from these mass functions, using co-moving volumes, as opposed to
physical volumes.  This tells us how the number of galaxies is evolving in
the same effective volume, with the effects of the Hubble expansion removed.
These mass functions are shown in Figure~1 down to the mass limit in
which they are complete based on all of these various surveys, which are
also listed in Table~1.

\section{The Evolution of Galaxy Number Density}

\subsection{Introduction and Caveats}

The primary method we use to determine the number densities of galaxies in the 
universe is to integrate through the fitted mass functions the number of 
galaxies which are at a given redshift.  This requires extrapolating fitted 
stellar mass functions to reach down to a low mass limit of the galaxy
population.  There are many ways in which this can be done which we
discuss below.  One of the most important questions is the lower limit 
at which we should count galaxies from the mass functions.
Due to the recent publication of stellar mass functions
up to $z \sim 8$ (e.g., Duncan et al. 2014; Grazian et al. 2015;
Song et al. 2015) we can now make this calculation for the first
time.   Another issue is whether or not the Schechter function can
be extrapolated below the limit of the data in which it was originally
fit. This is a question we investigate
in detail.
 
This is complementary to the directly observed
approach presented in the Appendix, and is a more accurate way to measure 
the number of galaxies in the currently observable universe, if 
the mass functions are properly measured and parameterized accurately. 
However this method is potentially fraught with 
pitfalls that have to be carefully considered and addressed.  
Not the least of which
is that this measurement relies on much more than simple photometry and
detection issues, which are also present when simply measuring the number
of galaxies.  The situation here involves other uncertainties
involving the measurements of stellar masses and redshifts.  Nonetheless, if
we can account for these uncertainties, the integration of fitted mass 
functions can tell us the number densities 
of galaxies within a given redshift interval with some measured
uncertainty.

We use this method to calculate the total number density of galaxies 
which are within the currently observable universe as a function of redshift. 
To do this we do not directly integrate the observed mass functions, but
 use a parameterized fitted form as given by the Schechter (1976) function
to determine the total number densities of galaxies as a function of
redshift.   The form of this function is given by:

\begin{equation}
\phi (M) = b \times \phi^{*} \ln(10) [10^{b(M - M^{*})}]^{(1 + \alpha)} \mathrm{exp} [-10^{b(M - M^{*})}]
\end{equation}

\noindent where $b=1$ for the mass function and $b=0.4$ for the 
luminosity function, which would be written in terms of absolute magnitudes.
For the mass function, M$^{*}$ is the characteristic mass in log units 
and determines where the mass function changes slope,
and M = $\log$(M$_{*}$/ \solm), is the mass in log units. Similarly for the 
luminosity function, M$^{*}$ corresponds to the characteristic magnitude.
For both functions $\phi^{*}$ is the 
normalization, and $\alpha$ determines the slope for fainter and lower
mass galaxies.  Our method is to use published values of 
$\phi^{*}$, $\alpha$ and M$^{*}$ to  
calculate the integrated number of galaxies within different 
redshift bins. 

We use the Schechter function as a tool to calculate the total number density 
as overall it does a good job of representing the distribution of galaxy 
masses at all redshifts in the ranges where we probe.  We do not however know at
what lower mass limit it remains valid, which is one uncertainty 
in our analysis.  We discuss the use of a M$_{*} > 10^{6}$ \solm limit below 
and the justification for using this as our lower limit. We also discuss 
how our results would change if we were to use a different lower mass cut 
off limit.

\begin{deluxetable*}{llllcl}
\tablecaption{Schechter Function Parameters \label{tab:table}}
\tablehead{
\colhead{Redshift (z)} & \colhead{$\alpha$} & \colhead{log $M^{*}$} & \colhead{$\phi^{*}$ ($\times 10^{-4}$)} & Limit & \colhead{Reference} \\
\colhead{} & \colhead{} & \colhead{\solm} & \colhead{Mpc$^{-3}$} & log M$_{*}$ & \colhead{}
}
\colnumbers
\startdata
0.20-0.40  & -1.19$\pm$0.08 & 11.20$\pm$0.10 & 22.4$\pm$6.0 & 8.0 & Perez-Gonzalez+08 \\
0.20-0.50  & -1.29$\pm$0.01 &  11.44$\pm$0.03  & 12.2$\pm$0.5 & 8.0  & Muzzin+13 \\
0.20-0.50  & -1.35$\pm$0.04 & 11.27$\pm$0.10 & 10.9$\pm$2.8 & 8.0 & Tomczak+14 \\
0.20-0.70  & -1.11$\pm$0.10 &  11.22$^{+0.13}_{-0.12}$  & 18.2 & 9.2 & Fontana+04 \\
0.30-0.50  & -1.41$\pm$0.02 & 11.54$\pm$0.07 &  6.3$\pm$1.3 & 7.0 & Mortlock+13 \\
0.40-0.60  & -1.22$\pm$0.07 & 11.26$\pm$0.11 & 17.4$\pm$4.5 & 8.6  & Perez-Gonzalez+08 \\
0.40-0.60  & -1.22$\pm$0.02 & 11.23$\pm$0.03   & 14.3$\pm$1.0 & 7.5  & Fontana+06\tablenotemark{a} \\
0.40-0.70  & -1.14$^{+0.04}_{-0.04}$ & 11.15$\pm$0.06 & 18.3$\pm$2.4 & 8.5 & Pozzetti+07 \\
0.50-0.75  & -1.35$\pm$0.04 & 11.22$\pm$0.06 & 11.7$\pm$2.1 & 8.3 & Tomczak+14 \\
0.50-1.00  & -1.17$\pm$0.01 &  11.22$\pm$0.02  & 16.3$\pm$0.6  & 8.9 & Muzzin+13 \\
0.50-1.00  & -1.34$\pm$0.02 & 11.38$\pm$0.04 &  7.6$\pm$0.9  & 8.0 &  Mortlock+13 \\
0.50-1.00  & -1.21$^{+0.03}_{-0.02}$  & 11.31$^{+0.07}_{-0.08}$ & 18.6$\pm$2.4 & 8.5 & Kajisawa+09 \\
0.60-0.80  & -1.26$\pm$0.08 & 11.25$\pm$0.08 & 15.1$\pm$3.7  & 9.2 &  Perez-Gonzalez+08 \\
0.60-0.80  & -1.24$\pm$0.03 & 11.24$\pm$0.04 & 10.9$\pm$1.0 & 8.1 & Fontana+06\tablenotemark{a} \\
0.70-0.90  & -1.01$^{+0.07}_{-0.08}$ & 10.89$\pm$0.06  & 26.0$\pm$3.8 & 9.1 & Pozzetti+07 \\
0.70-1.00  & -1.27$\pm$0.10 &  11.37$^{+0.22}_{-0.21}$  & 11.0  & 10.4 & Fontana+04 \\
0.75-1.00 & -1.38$\pm$0.04 & 11.38$\pm$0.12 & 6.8$\pm$1.9  & 8.4 & Tomczak+14 \\
0.80-1.00 & -1.23$\pm$0.09 & 11.27$\pm$0.09 & 12.3$\pm$3.4   & 9.4 & Perez-Gonzalez+08 \\
0.80-1.00 & -1.25$\pm$0.03  &  11.26$\pm$0.05  & 8.5$\pm$0.9 & 8.2 & Fontana+06\tablenotemark{a} \\
0.90-1.20 & -1.10$^{+0.07}_{-0.08}$ & 11.00$\pm$0.06 & 18.3$\pm$2.8 & 9.2 & Pozzetti+07 \\
1.00-1.25 & -1.33$\pm$0.05 & 11.31$\pm$0.10 & 6.5$\pm$1.9  & 8.7 & Tomczak+14 \\
1.00-1.30 & -1.26$\pm$0.04 & 11.31$\pm$0.11 & 8.7$\pm$2.0 & 9.4 & Perez-Gonzalez+08 \\
1.00-1.40 & -1.28$\pm$0.04 &  11.26$\pm$0.07  & 6.2$\pm$0.8  & 8.3 & Fontana+06\tablenotemark{a} \\
1.00-1.50 & -1.36$\pm$0.05 & 11.43  & 6.0$\pm$1.1 &   8.6 & Mortlock+11 \\
1.00-1.50 & -1.32$^{+0.04}_{-0.04}$  & 11.36$^{+0.13}_{-0.10}$ & 6.9$\pm$1.4 & 9.0 & Kajisawa+09 \\
1.00-1.50 & -1.31$\pm$0.03 & 11.26$\pm$0.04 &  6.2$\pm$0.9 & 8.5 & Mortlock+13 \\
1.20-1.60 & -1.15$^{+0.12}_{-0.12}$ & 10.94$\pm$0.07 & 14.8$\pm$3.0 & 9.8 & Pozzetti+07 \\
1.25-1.50 & -1.29$\pm$0.05 & 11.10$\pm$0.05 & 7.8$\pm$1.6  & 8.8 & Tomczak+14 \\
1.30-1.60 & -1.29$\pm$0.08 & 11.34$\pm$0.10 & 5.4$\pm$2.0 & 9.8 &  Perez-Gonzalez+08 \\	
1.40-1.80 & -1.31$\pm$0.06 & 11.25$\pm$0.11  &  4.3$\pm$0.7 & 8.5  & Fontana+06\tablenotemark{a} \\
1.50-2.00 & -1.51$\pm$0.03 & 11.37$\pm$0.06 &  1.8$\pm$0.4 & 8.5 & Mortlock+13 \\
1.50-2.00 & -1.19$\pm$0.06 & 11.43   & 7.5$\pm$1.2 & 9.3 & Mortlock+11\tablenotemark{b} \\
1.50-2.00 & -1.33$\pm$0.05 & 11.25$\pm$0.05 & 5.2$\pm$1.1 & 9.0 & Tomczak+14 \\
1.50-2.50 & -1.45$^{+0.06}_{-0.06}$ & 11.32$^{+0.13}_{-0.10}$ & 3.1$\pm$0.7  & 9.3 & Kajisawa+09 \\
1.80-2.20 & -1.34$\pm$0.07 & 11.22$\pm$0.14  &  3.1$\pm$0.6   & 8.7 & Fontana+06\tablenotemark{a} \\
2.00-2.50  & -1.56$\pm$0.06 & 11.24$\pm$0.10 &  1.7$\pm$0.4  & 9.0 & Mortlock+13 \\
2.00-2.50  & -1.50$\pm$0.08 & 11.43  & 3.5$\pm$0.9 & 9.4 & Mortlock+11\tablenotemark{b} \\
2.00-2.50  & -1.43$\pm$0.08 & 11.35$\pm$0.13 & 2.6$\pm$0.9  & 9.3 & Tomczak+14 \\
2.20-2.60  & -1.38$\pm$0.08 & 11.16$\pm$0.18  &  2.4$\pm$0.5   & 9.0 & Fontana+06\tablenotemark{a} \\
2.50-3.00  & -1.69$\pm$0.06 & 11.26$\pm$0.12 &  0.9$\pm$0.4 & 9.3 & Mortlock+13 \\
2.50-3.00  & -1.74$\pm$0.12 & 11.57$\pm$0.33 & 0.4$\pm$0.4  & 9.5 & Tomczak+14 \\
2.50-3.50  & -1.59$^{+0.13}_{-0.14}$  & 11.39$^{+0.32}_{-0.20}$ & 1.0$\pm$0.5  & 9.5 & Kajisawa+09 \\
2.60-3.00  & -1.41$\pm$0.09 & 11.09$\pm$0.21  &  1.9$\pm$0.4  & 9.2 & Fontana+06\tablenotemark{a} \\
3.00-3.50  & -1.45$\pm$0.11 & 10.97$\pm$0.27  &  1.5$\pm$0.3  & 9.4 & Fontana+06\tablenotemark{a} \\
3.00-3.50  & -1.86$^{+0.05}_{-0.04}$ & 11.45$\pm$0.07 & 0.4$\pm$0.2  & 10.4 & Caputi+11 \\
3.50-4.00  & -1.49$\pm$0.12 & 10.81$\pm$0.32  &  1.1$\pm$0.3  & 9.6 & Fontana+06\tablenotemark{a} \\
3.50-4.25  & -2.07$^{+0.08}_{-0.07}$ & 11.37$\pm$0.06 & 0.1$\pm$0.1 & 10.4 & Caputi+11 \\
3.50-4.50  & -1.53$^{+0.07}_{-0.06}$ &  10.44$^{+0.19}_{-0.18}$ &  3.0$^{+1.8}_{-1.2}$ & 8.5 & Song+15 \\
3.50-4.50  & -1.89$^{+0.15}_{-0.13}$  & 10.73$^{+0.36}_{-0.32}$ &  1.9$^{+3.5}_{-1.3}$ & 8.5 & Duncan+14 \\
3.50-4.50  & -1.63$\pm$0.09 &  10.96$\pm$0.18 &  1.2$\pm$0.4  & 8.5 & Grazian+15 \\
4.25-5.00 & -1.85$^{+0.27}_{-0.32}$ & 11.06$\pm$0.1 & 0.1$\pm$0.1 & 10.5 & Caputi+11 \\
4.50-5.50  & -1.74$^{+0.41}_{-0.29}$ &  10.90$^{+0.98}_{-0.46}$ &  1.2$^{+4.8}_{-1.2}$  & 8.7 & Duncan+14 \\
4.50-5.50  & -1.67$^{+0.08}_{-0.07}$ &  10.47$^{+0.20}_{-0.20}$ &  1.3$^{+1.0}_{-0.6}$ & 8.7  & Song+15 \\
4.50-5.50  & -1.63$\pm$0.09 &  10.78$\pm$0.23 &  0.7$\pm$0.3  & 8.7 & Grazian+15 \\
5.50-6.50  & -2.00$^{+0.57}_{-0.40}$ &  11.09$^{+1.13}_{-1.06}$ &  0.1$^{+4.1}_{-0.1}$  & 9.0 & Duncan+14 \\
5.50-6.50  & -1.55$\pm$0.19 &  10.49$\pm$0.32 &  0.7$\pm$0.5  & 8.9 & Grazian+15 \\
5.50-6.50  & -1.93$^{+0.09}_{-0.09}$ &  10.30$^{+0.14}_{-0.15}$ &  0.3$^{+0.3}_{-0.1}$  & 9.0 & Song+15 \\
6.50-7.50  & -2.05$^{+0.17}_{-0.17}$ &  10.42$^{+0.19}_{-0.18}$ &  0.1$^{+0.1}_{-0.5}$  & 9.2 & Song+15 \\
6.50-7.50  & -1.88$\pm$0.36          &  10.69$\pm$1.58          &  0.1$\pm$0.1  & 9.2 & Grazian+15 \\
\enddata
\tablenotetext{a}{Fontana et al. (2006) fit the evolution of the Schechter parameters over redshift and not within individual redshift bins.  These values are thus derived by using their fitting formula for the parameters and their associated errors.}
\tablenotetext{b}{Mortlock et al. (2011) use a constant M$^{*}$ at log M$^{*}$ = 11.43 across the redshift range they study from z = 1 to 3. }
\tablecomments{This table lists the parameters of the fitted Schechter functions
which we use to carry out our calculations.  These fits are all normalized to
have the same Salpeter IMF with  Pozzetti et al. (2007), Duncan et al. (2014),
and Mortlock et al. (2015) originally using a Chabrier IMF, and 
Muzzin et al. (2013) using a Kroupa IMF.}
\end{deluxetable*}

Because we are integrating these mass functions throughout the universe's
history we must use a variety of surveys to account for the number
of galaxies at different redshifts.  Different redshift ranges require
surveys done at different wavelengths, and various surveys have found
sometimes differing values for the Schechter parameters.  In this paper
we attempt a comprehensive examination of these mass functions
which, particularly at low redshift, can give widely divergent number densities
and forms of evolution.   We find nearly identical results if we use the
double forms of Schechter fits sometimes presented for the fits to mass functions at lower redshifts, or if we use power-law fits to the mass functions for the highest redshift galaxies.

Between redshifts $z \sim 0-3$ we use the fitted mass function parameter 
values and their errors from surveys carried out by
Perez-Gonzalez et al. (2008), Kajisawa et al. (2009), Fontanta et al.
(2004, 2006), Caputi et al. (2011), Pozzetti et al. (2007), 
Mortlock et al. (2011), and Mortlock et al. (2015).     These are stellar mass 
functions are determined by measuring the stellar masses of objects via
SED fitting.     While there is a large scatter in the various measurements of
the Schechter function parameters, we use all this information to take into
account different measurement and fitting methods as well as cosmic variance from
the different fields used.   These mass functions, as parameterized by the
Schechter function, are shown in Figure~1.  We also convert those studies 
that use a Chabrier IMF -
which are: Pozzetti et al. (2007), Duncan et al. (2014), Mortlock et al. 
(2015) and Muzzin et al. (2013) who uses a Kroupa IMF into a Salpeter IMF. 
 The list of values we use in our analysis is shown in Table~1.

Note that we only consider those
mass functions where the parameter $\alpha$ in the Schechter fit was allowed
to vary.  If a mass function result is obtained from 
fixing the value of $\alpha$ this biases the number of galaxies, as this 
value has a major impact on the number of fainter, and lower mass, 
galaxies in given volume (\S 3.2). We therefore exclude mass function results 
from studies who hold $\alpha$ 
constant when fitting for the other Schechter parameters.  

Recently, the first measurements of the stellar mass function at high
redshifts up to $z \sim 8$ allows for this procedure to be carried out
back to when we can study the earliest galaxies discovered to date.  The 
mass functions we use come from Duncan et al. (2014) who use data from
the GOODS-S CANDELS field, 
Grazian et al. (2015) who use GOODS-S and GOODS-N
CANDELS data, and GOODS-S/N CANDELS and HUDF data from Song et al. (2015).

 At higher redshifts  mass functions are relatively new, thus we also
examine observed luminosity functions in the ultra-violet, typically at
1500\AA\, to test for consistency.  For this we use the luminosity functions 
published in Bouwens et al. (2011), McLure et al. (2009), 
McLure et al. (2013), Bouwens et al. (2015) and Finkelstein
et al. (2015). McLure et al. (2013) and Bouwens et al. (2015)
 discusses the results from the deepest HST data including the
HUDF12 survey, which probes
galaxies to the highest redshifts at $z = 8$ and $z = 9$.  

To convert our stellar mass limit to a UV magnitude limit we use the 
relations between
these two quantities as calculated in Duncan et al. (2014).  Duncan
et al. (2014) fit the linear relation between the mass and light in the UV
and how it evolves with redshift. We use these to determine the UV limit
corresponding to our default mass limit of M$_{*} = 10^{6}$ \solm.
This thereby permits us to associate our stellar mass limit with a
limit in absolute magnitude in the UV.  We do not
use these values in our calculations, but use these luminosity
functions to check consistency with our results from the
stellar mass functions.    We find a high consistency with the
stellar mass functions, including when
we use the different variations on the conversion between stellar mass
and UV luminosity (e.g., Duncan et al. 2014; Song et al. 2015).
Furthermore, our high-z mass functions all more
or less agree at high redshift, with the exception of Grazian et al. (2015)
whose results lead to a slightly lower value of $\phi_{\rm T}$.

\begin{figure}
 \vbox to 130mm{
\hspace{-1cm} 
\includegraphics[angle=0, width=90mm]{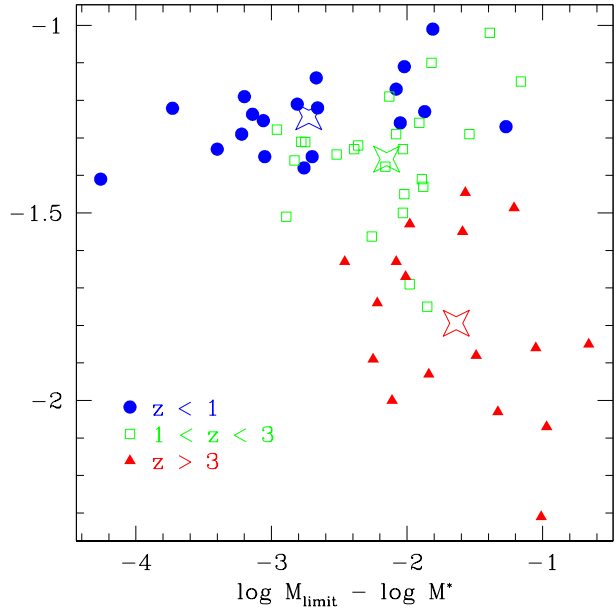}
\caption{The relationship between the fitted value of
$\alpha$ and the depth of the data in which the mass function was fit in
reference to the value of M$^{*}$ (see text).  Systems which are more
negative in log M$_{\rm limit}$ - log M$^{*}$ have deeper observed mass
functions with the deepest data at $z < 1$ probing a mass limit which is
a factor of 10$^{4}$ lower 
than M$^{*}$.  The colors of the points denote the redshift range, with the
blue circles those galaxies at $z < 1$, the green boxes those at
$1 < z < 3$ and the red triangles those at $z > 3$.  The large open
stars show the average values within each redshift range. }
\vspace{2cm}
} \label{sample-figure}
\end{figure}


\subsubsection{Integrating the Stellar Mass Function}

The major aspect of this paper is that we integrate the fitted Schechter
function for all of the mass function data below the limit in which the data
is obtained. There are two distinct questions concerning doing this. One is
if we are using fits from mass functions in which the data is not as deep as
others, then
are we technically able to retrieve the mass function correctly?  As the
number density depends strongly upon the value of the faint-end slope 
$\alpha$ (\S 3.2)
we can express this question another way:  If we have data down to some
limit, say M$^{*}$ and fit the Schechter parameters, will we retrieve the
same parameters if we were probing down to a low mass limit?  

The other
issue, we address in the next subsection has to do with how low we should 
integrate these mass functions.  Here we discuss the more technical point of
whether the Schechter function can be extrapolated at all with fits that
are based on a variety of mass functions of diverse depths.

\begin{figure}
 \vbox to 210mm{
\hspace{-0.25cm} 
\includegraphics[width=90mm, angle=0]{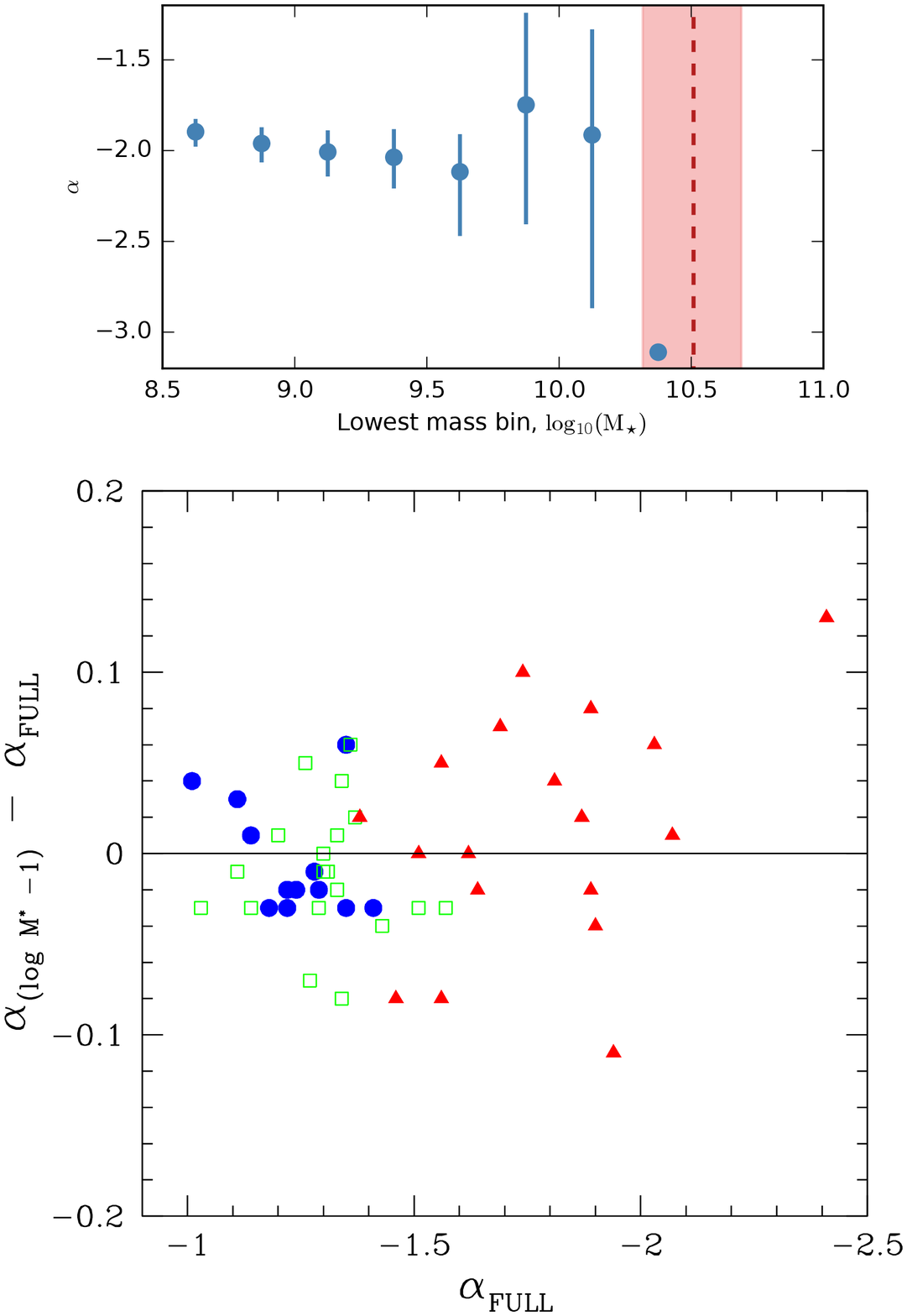}
 \caption{The two panels on this figure show our simulations and determinations
based on actual data for how the faint end slope $\alpha$ changes as a function of depth relative to the value of M$^{*}$.  In the top panel we show an example using the $z \sim 4$ mass function from Duncan et al. (2014) which reaches a nominal depth of log M$_{*} = 8.5$, which is 100 times deeper than the value of M$^{*}$ at this redshift.  The vertical dashed line shows this value of M$^{*}$, and
the shaded pink area shows the 1-sigma errors on the full M$^{*}$ fit,
which is the 16 and 84 percentiles in the marginalized distribution,
with the best-fit dashed line being the 50th percentile of this distribution.
Mathematically this is log M$^{*}$ = 10.50$^{+0.19}_{-0.17}$.  The blue points 
show the retrieved value of $\alpha$ when redoing the entire
Schechter function fits, but only to the quoted mass depth on the x-axis.  This
fitting is done through a full V$_{\rm max}$ methodology as outlined in
Duncan et al. (2014) and in the text.  The bottom panel shows the difference between the
fitted $\alpha$ values when using the full depth of the data ($\alpha_{\rm FULL}$), and when only fitting down to a mass limit of log M$^{*} - 1$ 
($\alpha_{\rm log M^{*} - 1}$).  Every mass
function we use in this paper is shown here.  The colours denote the redshift
range, as in Figure~2.   As can be seen, we are  able to retrieve
the values of $\alpha$ as long as we can reach at least 1/10 the value of
M$^{*}$.  The other Schechter function parameters are also accurately retrievable in these simulations. }
\vspace{5cm}
} \label{sample-figure}
\end{figure}

We address this issue in a few ways. We first show in Figure~1 the mass
functions we use as a function of redshift. The dashed, black lines,
show the extrapolation we use down to our M$_{*} = 10^{6}$ \solm limit. 
We show in Figure~2 the fitted $\alpha$ values for these mass functions
as a function of the depth of the data, parameterized by how deep the data
is in reference to the fitted M$^{*}$ values.  As can be seen, we do not
probe as deep at higher redshift.  We also see a trend, well established now
in several papers (e.g., Duncan et al. 2014; Song et al. 2015) that the value
of the fitted $\alpha$ becomes stepper at higher redshifts.   There is a
slight trend such that the fitted value of $\alpha$ becomes more steep (i.e.,
more lower mass galaxies) for deeper data, although this is not present at the
highest redshifts.

We also carry out two different simulations to demonstrate that we are
able to retrieve the correct value of $\alpha$, and the other Schechter
function parameters, M$^{*}$ and $\phi^{*}$, if we are probing about a factor
of ten less massive than M$^{*}$, or deeper.  Figure~3 shows the results of
these simulations.

The first one on the top panel of Figure~3 shows a full refitting of the lowest
redshift (z $\sim 4$) mass functions from Duncan et al. (2014).  These fits were
redone  using the emcee MCMC code (Foreman-Mackey et al. 2012).  The resulting fits thus technically differ from those in Duncan et al. (2014), 
 most significantly in the marginalized errors which are slightly smaller,
 but not in a significant way.  The dashed
vertical line shows the value of M$^{*}$ for this mass function when fit
at its deepest level down to the M$_{*} = 10^{8.5}$ \solm limit, which
is a factor of 100 times less massive than the value of M$^{*}$.
 The blue points show the value of the refitted $\alpha$
values of the Schechter function as a function of depth.  As can be seen,
within 0.5 dex of M$^{*}$ the mass function is difficult to fit correctly,
providing an incorrect $\alpha$ or one with a large error bar.  However
when the depth of the data is larger than 0.5 dex from M$^{*}$ we obtain
an accurate measure of $\alpha$.  

The bottom plot of Figure~3 shows a simulation whereby all of the mass 
functions we use were resampled to end at 1/10th the value of M$^{*}$ (1 dex
in log space) and
then fit with the resulting faint end slope, $\alpha_{\rm log M^{*} - 1}$.  
As shown in Figure~2, all of our mass functions are complete to 1/10 the value
of M$^{*}$ or deeper. 
These simulated mass functions take
into account the number of galaxies present in each bin, and therefore also
the statistical uncertainty when dealing with smaller numbers of galaxies.

We then compare this simulated value 
with the full mass range fitted value $\alpha_{\rm FULL}$ which probes
to the depths shown in Figure~2. 
This difference 
is plotted as a function of the full value $\alpha_{\rm FULL}$ in
Figure~3.  This shows that as a function of redshift, and shape
of the mass function as parameterized by $\alpha_{\rm FULL}$, the value of
$\alpha$ can be easily retrieved when the depth of the mass function is
at least 1/10th the value of M$^{*}$.  This hold as well for the other
Schechter function parameters.  We therefore use only those mass functions
which are at least a factor of ten deeper than their fitted M$^{*}$ value
(Figure~2).

 We must also consider the effects of the covariances between 
M$^{*}$, $\phi_{\rm T}$, and $\alpha$ when these parameters are fit together.
The resulting values of these parameters  are 
correlated, as can be seen in their distributions from various mass 
function fits (e.g., Fontana et al. 2006; Grazian et al. 2015; Weigel et al.
2016).  For example, the 
best-fitting values of $\alpha$ can be strongly affected by a 
few galaxies at the massive end of the
mass function, which should in principal have little effect on the 
low-mass slope, but when fitting can.  This covariance is such that there is a strong correlation
between the values of $\alpha$, $\phi^{*}$ and M$^{*}$ such that the value of
$\alpha$ becomes more negative as M$^{*}$ increases, and $\phi^{*}$ goes up when
M$^{*}$ does down.  This is such that a higher fitted
M$^{*}$ increases the resulting value of $\phi_{\rm T}$, and a more negative 
$\alpha$ does so as well, therefore the covariance between these 
is a potentially serious issue.  However, as there is a corresponding drop in 
the value of 
$\phi^{*}$ when M$^{*}$ increases, this negates to some extent the
strong covariance between $\alpha$ and M$^{*}$.

The mass function fitting we use does not often provide the covariance
between these parameters in the respective results we use.  The error 
bars on the fits however do represent a good
representation of the extent of the range of $\phi_{\rm T}$, when combined together.
That is, the error on $\phi_{\rm T}$ is calculated by taking the extent of the 
error-bar when $\alpha$ is at its maximum negative extent, M$^{*}$ is at its
highest value and $\phi^{*}$ is at its lowest value for an upper error limit, 
and vice-versa for the lower error limit.

We investigate the total effect of this covariance by simulating the covariance
for the $z=4$ mass function from Grazian et al. (2015).  We show the distribution 
of $\phi_{\rm T}$ values when we randomly sample 1000 times across the 
covariance of the
Schechter function fitting parameters in Figure~4.  The distribution of 
$\phi_{\rm T}$ is not Gaussian, but has a tail at higher values of 
$\phi_{\rm T}$. However, the average value of this distribution, 
$\phi_{\rm T} = 0.27$, which is just slightly higher than the 
value calculated with the best fitted Schechter
values which is  $\phi_{\rm T} = 0.24$.   The error range calculated with
our method is also similar to the distribution of these points. 
Other covariance analyses results in similar conclusions.

\begin{figure}
 \vbox to 130mm{
\hspace{-1cm} 
\includegraphics[angle=0, width=90mm]{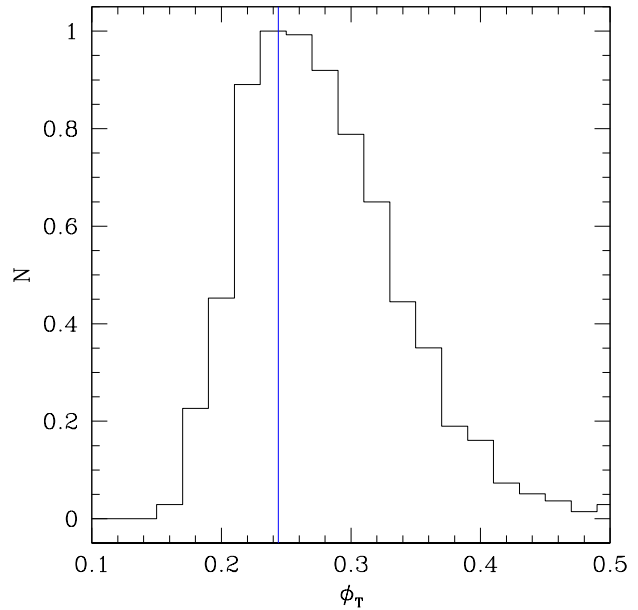}
\caption{The relative distribution of $\phi_{\rm T}$ values derived from the
sampling of the co-variance between the parameters of $\alpha$,
$\phi^{*}$ and M$^{*}$ for one of the mass functions used
at $z = 4$ from Grazian et al. (2015).  The average value of
the distribution is $\phi_{\rm T}$ = 0.27 which is close to the
derived value of $\phi_{\rm T}$ = 0.24 which is shown by the vertical
blue line.  The range here is also similar to the resulting errors we
calculate for $\phi_{\rm T}$. }
\vspace{2cm}
} \label{sample-figure}
\end{figure}

\subsubsection{The Galaxy Stellar Mass Lower Limit}

Another main issue with these mass functions is how deep to integrate
to obtain the total number density of galaxies at a given redshift.  Our choice
here is important, as the total number can depend significantly on the limits
used in the integration.  Our initial and fiducial low-mass limit is  
M$_{*} = 10^{6}$ \solm, which is the typical lower limit for dwarf galaxies 
in the nearby universe.  Most systems at masses lower than this are star 
clusters within galaxies, or debris from galaxy interactions.  For example, 
many of the galaxies in the Local Group and nearby clusters with masses 
lower than this are likely either of tidal origin, or are 
misidentified star clusters (e.g., Ibata et al. 2013; 
Penny \& Conselice 2008). 

While there are certainly some galaxies at  
lower masses than this in the local universe, by using a M$_{*} 
= 10^{6}$ \solm limit we limit our exposure
to the risk of extrapolating the mass function too deeply. Because of the use
of a limit whereby we are probing beyond the direct observations,
 we cannot be 100\% certain
that the mass function is necessarily valid to these depths.  Also, the 
exact value of this lower limit will drive the quantitative 
values of our total number density strongly.
Therefore the nominal total number density we calculate is based on the number
of galaxies greater than M$_{*} = 10^{6}$ \solm.  We also assume that the 
mass functions of high redshift galaxies do not change slope at lower mass
limits than we currently probe. If they do so, then our results would be 
different. However, for the reasons given above, and below, this 
change in slope is unlikely as this change in mass function shape at low
masses has not seen at 
any redshift including at lower redshift where we can probe deeper than at 
higher redshift.

We also justify using these mass limits, and thus extrapolating our mass 
functions down to M$_{*} = 10^{6}$ \solm, in other ways.  Whilst this paper has
nothing directly to say about reionization, many authors have shown that 
an extrapolation of the mass function to at least this mass limit is needed
 to reionize the universe. As shown in papers such as Robertson et al. (2015),
Duncan \& Conselice (2015) and Dufy et al. (2014), the integration
of the measured mass function at $z \sim 7$ to at least this M$_{*} =
10^{6}$ \solm limit is necessary and sufficient to reionize the universe.   
While this does not imply that these galaxies are reionizing the 
universe, it does show
that this limit is required if UV emission from galaxies is the ionization culprit.  
This is also consistent with recent determinations of the cosmic background 
light which show that there are a factor of 10 more galaxies than we can 
presently observe, per unit area, needed to account for this light  
(e.g., Mitchell-Wynne et al. 2015)

Furthermore, predictions of Cold Dark Matter show
that the luminosity and mass functions of galaxies should be very steep
with a mass function slope of $\alpha \sim -2$ (e.g, Jenkins et al. 2001), 
it is thus reasonable to assume
that measured luminosity and mass functions continue to increase past
the arbitrary observational limit we can reach with observations today.
Furthermore these theoretical models show that there are indeed galaxies down 
to these low masses, both through dark matter halos and through simulated
galaxies (e.g., Jenkins et al. 2001; 
Gonzalez-Perez et al. 2014; Henriques et al. 2015).

Furthermore, the stellar populations of low mass galaxies in the local
universe show that many of the stars in these galaxies were formed at very 
early times, and thus were present at these high redshifts (e.g., Grebel, 
Gallagher \& Harbeck 2003), and that by examining the mass function of Local 
Group galaxies we also obtain a high number of low mass systems with a 
retrieved steep $\alpha$ value (e.g., Graus et al. 2015).  We however 
later test how 
our results would change if we were 
to use a different observational limit to constrain this uncertainty.

\subsection{Integrated Number Density Evolution}

We integrate the fitted Schechter parameters for the various fits to the
mass functions given in the papers described in the 
previous section to obtain a measurement of
the total number densities of galaxies.   

We calculate at a given redshift bin
the total number density of galaxies ($\phi_{\rm T}$), 
given the parameters of the Schechter
fit to the mass function, $\phi(M,z)$:

\begin{equation}
\phi_{\rm T} (z) = \int_{M_{1}}^{M_{2}} \phi (M, z)\, {\rm dM}
\end{equation}

\begin{figure*}
\vspace{-1cm} 
 \vbox to 130mm{
\includegraphics[angle=0, width=170mm]{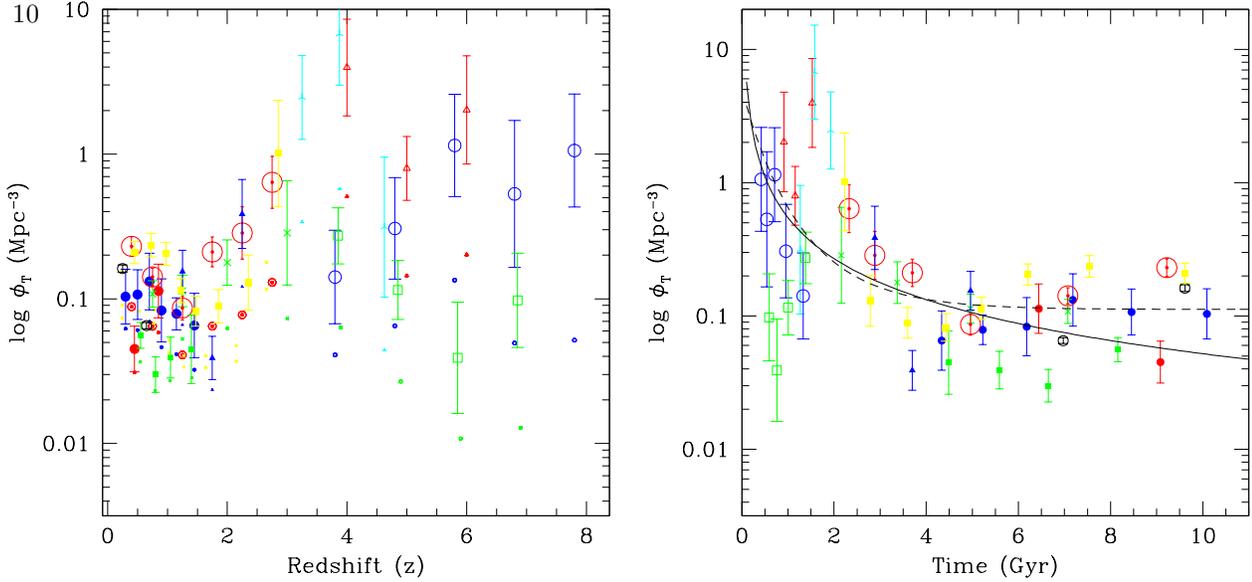}
 \caption{\small Figure showing the evolution of the co-moving total number 
densities for galaxies within the universe as a function of both redshift 
and time.  Shown
here are the various surveys which are used in these measurements.  This
includes measurements from Perez-Gonzalez et al. (2006) (blue solid circles 
at $z < 2$); Fontana et al. (2004) (large solid red circles at $z < 1$); 
Kajisawa 
et al. (2009) (green crosses at $z < 4$); Mortlock et al. (2015) (large red
open circles at $z < 4$); Muzzin et al. (2013) (black squares at $z < 1$);  
Pozzetti et al. (2007) (green boxes at $z < 2$); 
Mortlock et al. (2011) (blue triangles at $z < 4$); Caputi et al. (2011) (cyan
up arrows at $z \sim 4$)  and
Tomczak et al. (2012) (solid yellow squares at $z < 3$).  For the 
higher redshift points we use the mass functions from Duncan et al. (2014) (red triangles at $z > 4$); Song et al. (2015) (blue open circles
at $z > 4$) and Grazian et al. (2015) (green open boxes at $z > 4$).
The smaller symbols
without error bars, usually towards the bottom of the plot on the left 
panel, show the integrated values for
each mass function to the completeness limit reached by
each survey.   The solid
line in the right panel is the best fits to the relation of $\phi_{\rm T}$ with
time, while the
dashed line  is the merger model best fit discussed
in \S 4.3.     }
} \label{sample-figure}
\end{figure*}

\begin{figure}
 \vbox to 120mm{
\hspace{-0.25cm}
 \includegraphics[angle=0, width=90mm]{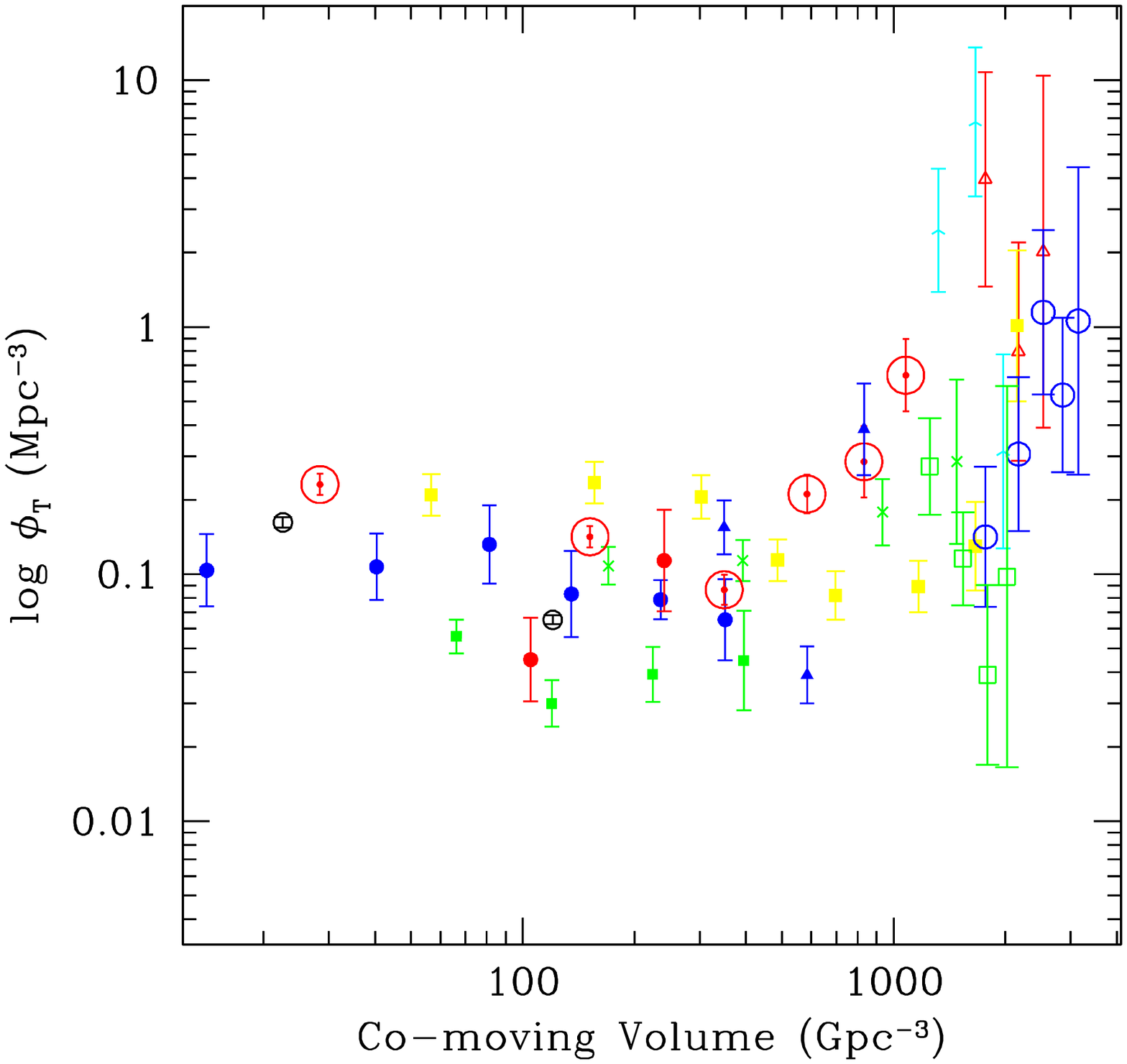}
\caption{Similar to Figure~5 but showing the change in total galaxy number
density as a function of total co-moving volume out to a given redshift.  
This total galaxy 
density increases with 
larger volume as Vol$^{1.28\pm0.38}$. The  symbols are the same as Figure~5.   This figure shows that the majority of the galaxies in the currently observable universe are at high redshift. }
\vspace{-1.5cm}
} \label{sample-figure}
\end{figure}

\noindent  A good approximation for this integral into an analytical formula can be obtained by taking the approximation for the Schechter function when the 
stellar mass limit is low.  The formula for this is given by:

\begin{equation}
\phi_{\rm T} (z) \approx \frac{-\phi^{*} 10^{(\alpha + 1)(M_{2} - M^{*})}}{(\alpha + 1)},
\end{equation}

\noindent showing that the number density is a function of all the Schechter
function parameters, but is especially dependent on the value of $\alpha$,
and the lower limit to integration $M_{2}$.  However throughout this paper
we do not use this approximation, but carry out direct numerical integration
of eq. 2.

When we plot the integrated number density $\phi_{\rm T}$ 
(Figure~5) for each data point from our various studies we find
some difference, at the same redshifts, between the various surveys for calculating the
total number density from M$_{1} = 10^{6}$ \solm to M$_{2} = 10^{12}$ \solm.  
We find a general increase with redshift in the integrated number density, 
although
some studies, such as Perez-Gonzalez et al. (2008) find a slight decline
from $z=0$ to $z=4$.   
On the other hand, studies that are based on 
very deep NIR imaging, such as Kajisawa et al. (2009), Caputi et al. (2011)
and Mortlock et al. (2011, 2015)
find an increase with redshift in $\phi_{\rm T}$ with higher redshift.     
This is due to the fact that the fitted $\alpha$ values are steeper at higher
redshifts in these studies.   More
recent results also find a steep $\alpha$ at high redshifts, which provides
a large value of $\phi_{\rm T}$, including Tomczak et al. (2012),
Duncan et al. (2014), Grazian et al. (2015) and
Song et al. (2015).    

 A steep $\alpha$ values continues to be found
in preliminary mass functions of even deeper studies utilizing lensing
from the Hubble Frontier Fields (Laporte et al. 2014) and the deepest
HST field data, with $\alpha$ values approaching or exceeding
$\alpha = -2$ (e.g., Bouwens et al. 2015; Livermore et al. 2016).  Thus, 
there are
more low mass galaxies per massive galaxy at high redshifts than in the 
local universe.  

We summarize these results and their errors in Figure~5.  Note that the 
integration of the quantity given in eq. 2 and plotted here, 
shows that while there are
generally more galaxies at higher redshift, the exact number has some 
variation.   From Figure~5 it is however clear that there is a rapid increase
in $\phi_{\rm T}$ up to $z = 3$ with a more mild increase at higher redshifts.

We also show in Figure~5 on the left panel the number densities 
of galaxies down to the
limit of integration which matches the completeness of each point in the
original data.  This is shown as the small points just below the main
data points which have error bars. This shows that our extrapolations 
to the lowest mass
limits is largest for the highest redshift galaxies by sometimes a factor
of 10 or so, and much less at lower redshifts.




When we fit the relation of $\phi_{\rm T}$ as a function of time we find that 
the relationship between the total
number density, $\phi_{\rm T}$ and time ($t$) is such that the increase
is slower at lower redshift $z < 1$ than at higher redshift.  
This shows that there is a rapid increase in the number of
galaxies in the universe above our nominal threshold of
M$_{*} = 10^{6}$ \solm.  The formal fit with time is

\begin{equation}
{\rm log} \phi_{T} (t) = (-1.08\pm0.20) \times {\rm log} (t) - 0.26\pm0.06 ,
\end{equation}

\noindent where $t$ is the age of the universe in units of Gyr.  
This equation shows that the number
of galaxies in the universe declines with time, with an overall decrease which
goes as $\sim 1/t$.  We find that a fit with time can be parameterized
well in this way, although a fit with redshift does not produce a single
well fit function.  We also show in Figure~6 the change in the number 
densities of galaxies as function of co-moving volume, showing that the 
majority of the galaxies in the universe are in the early universe.  This 
also shows likewise that the average total number density of galaxies 
declines as time goes on.    

We also show in Figure~7 the number density evolution at
higher mass limits of M$_{*} = 10^{7}$ \solm, and  at
M$_{*} = 10^{10}$ \solm.  What we find when examining these results at different
masses is a few things. First it appears that there is a turn over in the
evolution of $\phi_{\rm T}$, such that for massive galaxies there is a
step increase in the number density of galaxies above the limit of 
M$_{*} = 10^{10}$ \solm.  This is opposite to what we find when we examine
the evolution at the limit of M$_{*} = 10^{6}$ \solm.  This turn over occurs
at roughly M$_{*} = 10^{7}$ \solm, as shown in Figure~7.  This demonstrates
that at this limit the evolution of $\phi_{T}$ is nearly constant with 
redshift.

The implication for these results are discussed in the next section.  Overall,
we are now able to use these results to determine how many galaxies
are in the observable universe with our various mass limits, the likely methods by 
which galaxies are evolving in
number density, as well as the implications for background light in
the optical and near infrared based on these results.

\section{Implications}

The major result from this paper is that we now have a scientific measurement
of the evolution of the total galaxy number density up to $z = 8$.  
We investigate several implications for
these results, including the total number of galaxies in the universe,
galaxy evolution, extragalactic background light,
Olbers' paradox and future galaxy surveys.  

\subsection{Comparison to Theory}

One of the things we can do with this analysis of evolving number densities
is to compare our result to those of theory.  Recently, the Illustris 
simulation derived the number densities of
galaxies as a function of redshift (Torrey et al. 2015).  The Illustris 
simulation is a cosmological hydrodynamical simulation designed to probe 
galaxy formation processes using a periodic box of size L = 106.5 Mpc.  
The overall set up
and code is described in Vogelsberger et al. (2013), but in summary includes 
hydrodynamics, gravity, star formation with feedback, and radiative cooling.  
It does a  reasonable job of matching observations of the galaxy population 
(e.g., Genel 2014).

Here we compare directly with the predictions for the number of galaxies 
above a certain mass, or the cumulative number density of galaxies, which is
exactly what we have discussed earlier in this paper.  Torrey
et al. (2015) provide an analytical formula to compare our values with.  
However, up to our redshifts of interest, these are only provided down
to M$_{*}$ = 10$^{7}$ \solm using a Chabrier IMF.  We thus
compare our integration down to this mass limit with these models
in Figure~7.    We also convert our number densities to this IMF before comparing with the Torrey et al. (2015)
results.  
We also compare with a higher mass limit of M$_{*}$ = 10$^{10}$ \solm.
This comparison is shown in Figure~7 for these two mass limits.

What can be seen in these figures is that the cumulative number density
observed agrees reasonably well with the predictions for the highest mass
galaxies at M$_{*} > 10^{10}$ \solm, but does not do as well of a job at the 
lower mass limits.  The reason for this is that the low mass end of the mass
function, particularly at lower redshifts does not match the data, with the
theory values being too high.  Interestingly, this does not seem to be the
case at higher redshifts where the theory and data agree better.  The reason
for the disagreement at low redshift is likely due to the fact that the
data show increasingly shallow values of $\alpha$, the faint end slope,
compared to the simulation.   This comparison also shows that our
extrapolation down to low mass limit is not excessive as we are still
below what is predicted in theory.   Later in the next section
we discuss the likely cause of this
changing $\alpha$ value and why the total number of galaxies declines with
time.

\begin{figure*}
 \vbox to 120mm{
\includegraphics[angle=0, width=180mm]{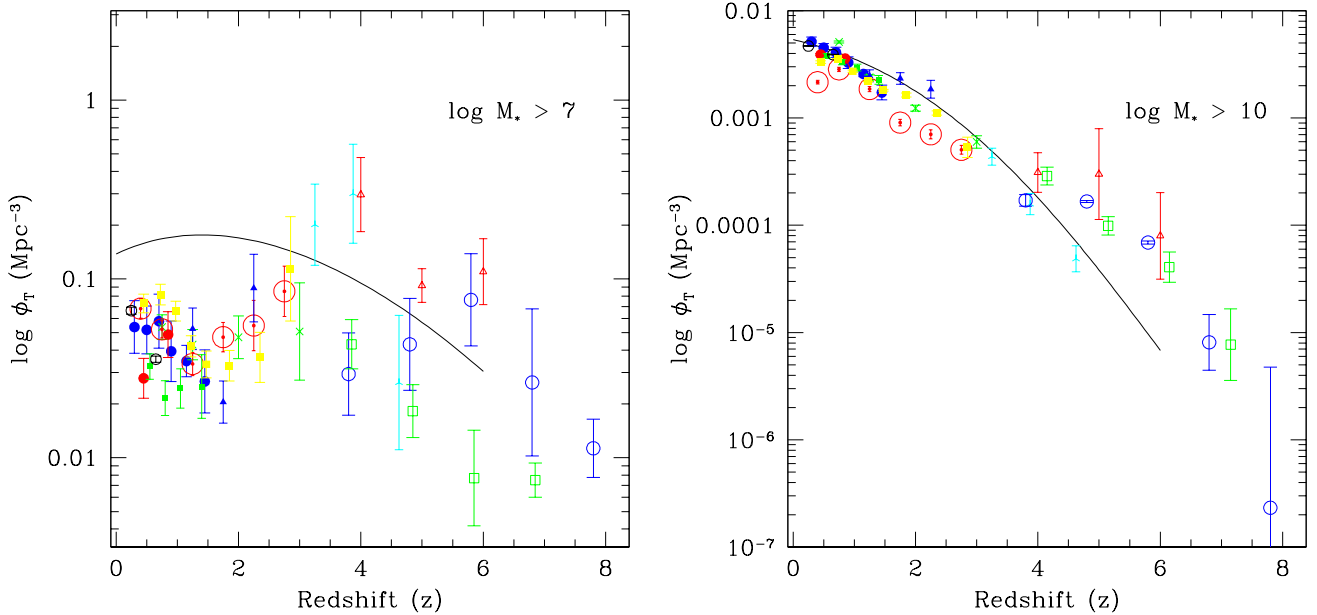}
 \caption{The number density evolution for galaxies selection with differing
mass limits than given on Figure~5.  In the left hand panel we show the
evolution of galaxies down to a limit of M$_{*}$ = 10$^{7}$ \solm, while
on the right hand side we show the evolution for a limit of 
M$_{*}$ = 10$^{10}$\solm.      The solid lines in both panels show the 
predicted number density evolution as parameterized by Torrey et al. 
(2015).  Note that the masses quoted here are those calculated with 
a Chabrier IMF to match the Torrey et al. (2015) models. The symbols and shapes
of the points denote the same survey as described in Figure~5.}
} \label{sample-figure}
\end{figure*}

\subsection{The Total Number of Galaxies in the $z < 8$ Universe}

The total number of galaxies in the universe is an interesting scientific
question, although it may not reveal anything fundamental about the cosmology
or underlying physics of the universe. None the less it is an interesting 
number  that should be known and quantified, although expressing it has
to be done within certain constraints.    

We use the results from \S 3 to investigate the total number of galaxies in the
currently observable universe down to a mass limit of
M$_{*} = 10^{6}$ \solm in two different ways.  We explore other limits that
give different answers later.  The first is through
simply taking the average total number of galaxies $\phi_{\rm T}$ at a given
redshift in Figure~5 and calculating its scatter.  When multiplied by
the volume at that redshift bin this gives us the average total number of
galaxies within that volume, which uses all the available data.   
Then we calculate the total number as simply the sum of these 
average numbers at each bin.  When we do this calculation we find that the 
total number of galaxies is given by,

$$N_{\rm tot, obs} = 2.8\pm0.6 \times 10^{12}$$ 

\noindent up to $z = 8$, which is almost a factor of ten larger than what is obtained from direct counts
from the deepest optical/NIR imaging data, as described in the Appendix.

To obtain another estimate of the total number of galaxies in the currently
visible Universe down to M$_{*} = 10^{6}$ \solm we integrate equation (4) 
from $0 \leq z \leq 8$ using 
our best fit relation. This avoids having to use a single measurement 
of the mass function to
obtain a result, and as such our measurement is based on all the available
data.  Because the mass functions we use are taken from a 
variety of fields, we are also accounting for cosmic variance within our
measurements.  To calculate the maximum and minimum limits to our total 
count, we take into account the errors assigned to $\phi^{*}$, $\alpha$ 
and M$^{*}$.
 
We use the fit above to calculate the total number of galaxies within
the currently visible universe down to M$_{*} = 10^{6}$ \solm from $z_1 = 0$ to
$z_2 = 8$, which is an integral over the number densities $\phi_{T} (z)$ and 
volume:

\begin{equation}
N_{\rm tot, fit} = \int_{t_{1}}^{t_{2}} \int_{0}^{4\pi}  D_{H} \frac{(1+z)^{2} D_{A}^{2}}{E(z)} \phi_{T} (t)\, d\,\Omega\, dt
\end{equation}

\begin{equation}
N_{\rm tot, fit} =(1.2^{+0.4}_{-0.2}) \times 10^{12}.
\end{equation}

\noindent where the volume is integrated over the entire sky through
$4\pi$ in steradians and $D_{A}$ is the angular size distance, $D_{H} 
= c/H_{0}$ and $E(z) = (\Omega_{M}\,(1+z)^{3} + \Omega_{\lambda})^{1/2}.$  
Thus knowledge of the redshift and the time since the Big-Bang are 
required to perform this integration.   

Performing this integral to $z = 8$ gives us 1.2 trillion individual galaxies, 
which
is again is just less than a factor of ten times higher than the number of 
galaxies which
in principle can be observed today in the universe with present
technology (see Appendix).    We use the difference between this 
value of the total number of galaxies and that calculated with the average
$\phi_{\rm T}$ values above to obtain a most likely value for the
average number of galaxies. As both methods are equally valid measurements
we take the average between them.   This gives us the ultimate final value 
of the total
number of galaxies within the currently visible universe down to M$_{*} = 
10^{6}$ ($N_{\rm tot, final}$),

\begin{equation}
N_{\rm tot, final} =(2.0^{+0.7}_{-0.6}) \times 10^{12}.
\end{equation}

\noindent This large value implies that there is a 
a vast number of galaxies that we have yet to discover at
magnitudes fainter than $m_{\rm max} \sim 29$ as we discuss in
Section \S 4.4.

However, as mentioned, this number depends strongly upon our use of a limit of 
M$_{*} = 10^{6}$ \solm, and our redshift limit of $z = 8$.  It is
important to explore how the total number would change if these limits
were different.   We find that the total
number does not depend much upon the redshift limit.  If we extend our
study to $z = 12$ by using the same number densities we find at $z = 8$,
our observational limit, then we find slightly more galaxies with the total
number increasing by 65\%.   There maybe very few galaxies at redshifts
higher than this.

In terms of the mass limit, there is more of a difference when using 
a lower limit.  While we have argued earlier that the mass
limit of M$_{*} = 10^{6}$ \solm is a natural one for galaxies,
it is important to discuss how the total number would change if we 
used a different stellar mass limit.  If we use a factor of ten 
smaller limit of M$_{*} = 10^{5}$ \solm, we find roughly a factor
of  $\sim 7$ more galaxies than using the 10$^{6}$ \solm limit. 
Thus our limit in equation 7 is
actually a lower limit in both terms of the mass limit and redshifts.  
The end result of this is that there are at least 2 $\times$
$10^{12}$ (two trillion) galaxies in the currently visible 
universe, the vast majority of  
which cannot be observed with present day technology as they are too faint.  

\subsection{Galaxy Evolution through Mergers and Accretion}

One of the major results we find is that the total number density of
galaxies in the universe declines with time from from high to low redshift 
when using a M$_{*} = 10^{6}$ \solm limit.  
In fact, what we find is that measured mass functions down to a limit 
of M$_{*} = 10^{6}$ \solm gives a total number density that 
declines by a factor of 10 within the first 
2 Gyr of the universe's history, and a further reduction at later times.  
This decline may further level off between redshifts of $z =1$ and 
$z = 2$.    The star formation
rate during this time is also very high for all galaxies, which should
in principle bring galaxies which wrer below our stellar mass limit 
into our sample at later times.   This would
naturally increase the number of galaxies over time, but we see the opposite.
  This is likely
due to merging and/or accretion of galaxies when they fall
into clusters which are later
destroyed through tidal effects, as no other method can reduce the number 
of galaxies above a given mass threshold.

We can use straightforward argument to demonstrate that hierarchical galaxy
formation must be occurring in the universe over time at this stellar mass
limit.  The first way we 
argue this is to calculate the number of galaxies in the nearby universe, 
and then infer the total 
number we would observe up to $z = 8$ if the co-moving number densities of 
galaxies remained the same.  We then compare this with the total number
of galaxies we can observe within the UDF-Max itself (see Appendix).  

Using the number
densities from the GAMA survey (Baldry et al. 2012), 
and by assuming a homogeneous universe, we find that there would be 
$\sim 2.8 \times 10^{11}$ galaxies at $z < 8$.  If
we use the inferred total number from eq. (7) then there are $\sim 7$ times
more observable galaxies than what we would predict in the universe if it had
the same number density as in the local universe.  There is likely no other
 way to remove these galaxies except through merging, destruction or accretion
with other systems.  


We investigate this further with a simple merger destruction model whereby 
when an individual merger or destruction event occurs the number of galaxies 
within a volume decreases by one.   Thus if there are $N_{\rm m}$ 
mergers per unit volume per unit time, then the number density declines 
by $N_{\rm m}$ galaxies in that volume, and over that time period.

We parameterize this overall merger rate 
in units of the number of mergers per unit volume per unit time as as
exponentially declining function:

\begin{equation}
R(t) = R_{0}\, {\rm e^{-t/\tau}}
\end{equation}

\noindent where $R_{0}$ the merger rate in units of number of mergers
per Gyr per Mpc$^{3}$ at $z = 0$, and $\tau$ characterizes
the global merger time-scale.   In this characterization
the number of mergers per unit volume which occur within a given redshift, 
or time, interval between time $t_1$ and $t_2$, which is equivalent to a
number density ($\phi_{\rm T}$) change, is given by

\begin{equation}
\phi_{\rm T}(t_{1}) - \phi_{\rm T}(t_{2}) = \int^{t_{2}}_{t_{1}} R_{0}\, e^{-t/\tau} {\rm d}t = R_{0}\, \tau ({\rm e^{-t_1/\tau}} - {\rm e^{-t_2/\tau}})
\end{equation}

\noindent if we measure this from $t_{1} = 0$, then the total integrated decline in
the number density $\phi_{\rm T}(t)$ is given by:

\begin{equation}
\phi_{\rm T} (t) = \phi_{\rm T}(0) - R_{0} \tau \times (1 - {\rm e}^{-t_2/\tau})
\end{equation}

\noindent where $\phi_{\rm T}(0)$ is the initial total number densities 
for galaxies at high redshift.

If we then use the observed data from Figure~5, and set $\phi_{\rm T}(z=8)= n_0 = 0.7$ 
Mpc$^{-3}$ then we find that the best fitting parameters of this 
merger model are  $R_0 = 1.28\pm0.20$ mergers Gyr$^{-1}$ Mpc$^{-3}$,
and $\tau = 1.29\pm0.35$ Gyr.    This model further predicts that at
$z = 1.5$ the merger rate is $R \sim 0.05$ mergers
Gyr$^{-1}$ Mpc$^{-3}$.  This compares well with the directly derived
merger rates from Conselice (2006) who find that the
merger rate for M$_{*} > 10^{8}$ \solm galaxies is $R \sim 0.03$ mergers
Gyr$^{-1}$ Mpc$^{-3}$, while for  M$_{*} > 10^{9}$  \solm galaxies the
rate is $R \sim 0.01$ mergers Gyr$^{-1}$ Mpc$^{-3}$.   These are also
close to the values found by Lotz et al. (2011) for more massive mergers
using an independent method, and to Bluck et al. (2012) who look at minor
merger pairs. 

One caveat of
this comparison however is that the merger rate we infer from the decrease in number densities in no way reveals what type of mergers these galaxies were involved with.  They could be in minor mergers with much more massive galaxies, 
or major mergers with galaxies of similar masses.  The comparison with 
Conselice (2006) is for a specific type of merger - major mergers with
mass ratios of 1:4 or higher.   However, the merger rate we derive 
from the evolution in number density is higher than the major merger rate
from Conselice (2006), and this is likely due to the increase presence of 
minor mergers for these systems. In fact, we find that the merger rate goes up
by about a factor of two for minor mergers (Bluck et al. 2012) at $z < 3$, 
suggesting that this is indeed the reason for the difference. Overall however
this argument demonstrates that a merger
interpretation is tenable given the similarity of merger rates 
derived using two very different ways to
measure the merger history.

Also, we find from Figure~7 that the number density evolution turns over for 
higher mass limits and in fact begins to increase with time for limits 
M$_{*} > 10^{7}$ \solm.  This implies that the majority of the merging occurs
at the lower mass range.  

The major conclusion from this is that galaxy merging and accretion/destruction
is the likely method
in which lower mass galaxies decline in number densities through 
cosmic time.  As 
discussed in e.g., Conselice (2014) this is also a major way in which 
galaxy formation occurs at these epochs. However, we have a limited amount
of information concerning the merging history at $z > 3$ (e.g., Conselice
\& Arnold 2009), yet some results show that the merger history may be
even more important at higher redshifts $1 < z < 3$ than at lower 
(e.g., Ownsworth et al. 2014) and this may extend to even higher redshifts. 

\subsection{Background light and Olbers' Paradox}

One of the implications for our results is that there should be a significant
number of galaxies in the Universe which with current technology we 
cannot observe, even within deep HST images. This naturally leads to 
the issue of background light in the universe in the observed optical, 
as well as other
problems such as Olbers' paradox, or scientifically understanding why the
sky is mostly dark at night.

\begin{figure}
 \vbox to 130mm{
\includegraphics[angle=0, width=90mm]{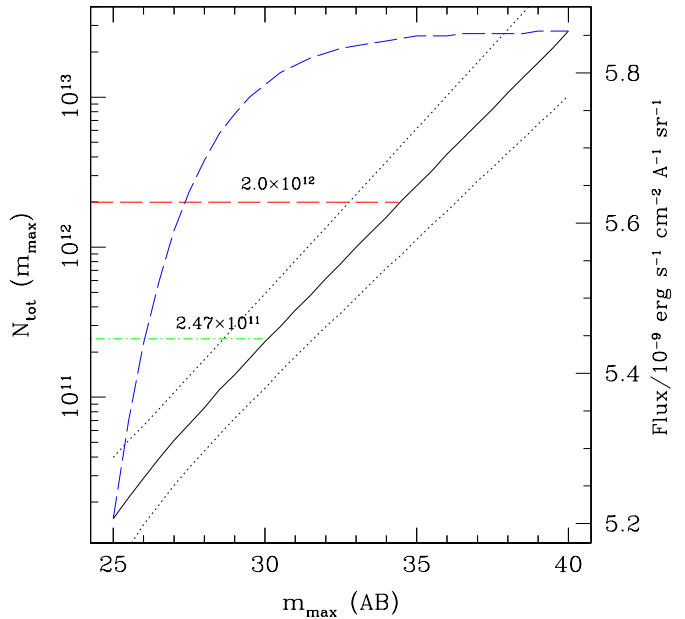}
 \caption{Plot showing the integrated number of galaxies as a function of
magnitude (solid black line and errors as dotted lines), and the integrated 
flux density of galaxies in
units of $10^{-9}\, {\rm erg~s^{-1} cm^{-2} \AA^{-1} sr^{-1}}$ as 
a function
of magnitude (blue dashed line).  The two dashed horizontal lines show
the observed total number of galaxies we can currently potentially observe
with present day technology (green dot-dashed) and the predicted total
number at $z < 8$ based on integrated stellar mass 
functions (red dashed).  }
} \label{sample-figure}
\end{figure}

We do not give a detailed scientific discussion of the background light
issue, as understanding the contributions of galaxies, as well as an
absolute measurement of the background light in the optical is
still controversial and difficult to measure 
(e.g., Bernstein 2007; Dominguez et al. 2011; Mitchell-Wynne et al. 2015).  
What we can do, using the 
results of this paper, is discuss how many galaxies
there are per unit area of the sky and what this implies about these background
light observations.

First, using our final value of N$_{\rm tot, final}$, we calculate that the 
projected density across all redshifts is on average 10,000 
galaxies per arcmin$^{2}$, 
which corresponds to an average of $\sim 3$ galaxies per arcsec$^{2}$. 
Using these numbers we can determine what 
fraction of the sky should be covered by galaxies.  If we take the 
hypothesis that these 3 per arcsec$^{2}$ galaxies 
occupy all the area on the sky, this 
gives an average size for these galaxies of 0.32 arcsec in radius.  This is
similar to the effective radii found for the faintest galaxies in deep 
HST imaging,
including the GOODS fields (e.g,. Ferguson et al. 2004; Bouwens et al. 2004),
as well as deeper imaging from HST WFC3 (Ono et al. 2013) for the highest
redshift systems. If the average galaxy is larger than 0.32 arcsec then it is
certain that every part of the sky is occupied by part of a galaxy
with stellar masses M$_{*} > 10^{6}$ \solm at $z < 8$.  
 Thus is appears that there are enough galaxies to
nearly cover the entire sky when using a 10$^{6}$ \solm limit.   If we were 
to use a lower integration limit such as 10$^{5}$ \solm the number of 
galaxies in the sky increases by a factor of $\sim 7$, making it nearly certain that all parts of the sky would be covered.

This demonstrates that, on average, every point in the sky should contain 
part of a galaxy in principle.   Most of these galaxies are however at 
higher redshifts, with the majority (around 2/3) at $z > 5$.  
With the corresponding
surface density of galaxies $\sim 0.8$ galaxies arcsec$^{-2}$ at
$z <5$.  Since most galaxies are 
extremely faint and cannot be easily observed with today's technology, 
we can only detect the presence of these galaxies through the cosmic 
background light (e.g., Bernstein 2007; Dominguez et al. 2011).  

Using the number of galaxies per magnitude and the magnitude limit that would
exist for a total number given in eq. (7) we calculate the total flux
density per steradians in the sky (Figure~8) as a function of magnitude.  We
do this by using the results in the Appendix, where we calculate the 
distribution of galaxy magnitudes in the UDF.  From this we calculate 
the total flux density
by integrating the flux density from each magnitude down to some magnitude
limit.

Figure~8 shows that at magnitudes around 30, and fainter, we find an 
integrated flux value $\sim$5.8$\times 10^{-9} 
{\rm erg s^{-1} cm^{-2} \AA^{-1} sr^{-1}}$. 
This value does not increase very much
at fainter magnitudes.  This flux compares well with the limits on the
optical background from e.g., Bernstein (2007) who finds an optical background
value in the I band of 7$\pm4 \times 10^{-9} {\rm erg s^{-1} cm^{-2} \AA^{-1} 
sr^{-1}}$.  Although the exact value of the background light
in the optical is still quite uncertain, this agreement with the best 
measurements thus far is encouraging.  Furthermore, we know that studies of 
Gamma-ray
Bursts show that the intrinsic star formation rate is a factor of 3-5 times
higher than that measured from deep imaging surveys (Ishida et al. 2011), 
demonstrating that there
are many faint galaxies yet to be detected at high redshifts.   Figure~8 
however shows that while the faint galaxies we have not
detected dominate the total number in the universe, they however do not 
contribute substantially to the unresolved background optical light beyond
$m \sim 32$.  

We can however use this to address the issue of why the 
sky is dark at night, a long standing problem
known as Olbers' paradox since Bondi (1952).  Our results 
reveal perhaps a new solution to this paradox.  There are various 
proposed solutions to Olbers' paradox but most solutions
can be placed into the one of two categories: (1) there are missing 
stars (and/or galaxies), or 
(2) the stars are there, but they cannot be observed for one reason or 
another.  Possible reasons for not detecting distant star light have included 
scenarios such as: absorption of light, hierarchical clustering, and the 
removal of energy from light in an expanding universe.    The traditional 
solution to the paradox is generally considered to be the fact that the 
universe is finite in age and size (Harrison 1987).

The human eye through L cone cells are sensitive out to a wavelength of
700nm.  If we consider which galaxies we would be able to see out to this 
wavelength, we find that galaxies out to $z \sim 5$ would still be detected
in principle.
Galaxies at higher redshifts would not be detected as their light is all 
shifted into wavelengths longer than 700nm or at the highest
redshifts would have their light blue-ward of Ly-$\alpha$ absorbed
by the Lyman-$\alpha$ forest.  At observed wavelengths bluer 
than this, all the light which is in principle detectable would originate 
within the UV continuum.  However all of this light is  absorbed by  
gas, and perhaps dust, within the host galaxies and in the
intergalactic medium near the galaxy itself, and along the site line to earth.

Thus, it would appear from this that absorption, long discarded as one 
objection to Olbers' paradox, is one method for removing objects from 
optical light detections.  This rest-frame UV light would however ionize 
hydrogen, which would then recombine, and in the process emit 
new light in the form of the Lyman-$\alpha$ line at 121.5 nm, and other lines
which are mostly redder than Lyman-$\alpha$  (e.g., Bertone
\& Schaye 2012).  However, this emission would be found mostly at observed 
near infrared wavelengths for systems at $z > 5$, outside 
the window of visible light detections.   The method of absorption 
thereby removes UV continuum light at wavelengths lower than the Lyman-limit,
and does not permit its re-introduction into the same, or bluer, 
wavelength range whereby it would be detectable within observed optical
light.  Furthermore, some of this light may also be absorbed
by dust, which in the limited time-range available, when the universe was 
only a few Gyr old at most, would then be re-emitted into the 
far-infrared, again outside the range of optical light.

It would thus appear that the solution to the strict interpretation of
Olbers' paradox, as an optical light detection problem, is a 
combination of nearly all possible solutions - redshifting effects, the
finite age and size of the universe, and through absorption. 

\subsection{Implications for Future Deep Surveys}

In this section we investigate future surveys and how the results of this
study have implications for future studies.  In the next 10 years we will
have telescopes that will image deeper than we currently can, with the ELTs
and JWST, and other telescopes and surveys that will cover vastly larger 
areas such as Euclid and LSST.  
 
First, we answer the question of when we would expect to find the observed
galaxy number counts to naturally turn over due to `running out' of
faint galaxies to detect.    While we are far from reaching
that limit today, if we assume that the slope of the number counts, 
as described in the appendix, remains
constant at magnitudes fainter than 29, we can determine at what 
magnitude limit
we will reach the total  number of actual galaxies.  We do this by integrating
the observed number of galaxies in the UDF using eq. (A3) and eq. (A4) 
as a function of magnitude down to limits ranging from m=20
to m = 40.  This integration allows us to infer the relationship between
the total number of galaxies and the limiting magnitude reached.
This relationship, shown in Figure~8, is given by:

\begin{equation}
{\rm log\, N_{tot}} = (4.97\pm0.03) + (0.21\pm0.01)\times {\rm m_{max}}
\end{equation}

\noindent We find that
the magnitude limit for the total number of galaxies calculated with
the mass functions (eq. 11) is $m_{\rm max}$ 
$\sim$ 34.9$^{+0.6}_{-0.8}$ AB mag, taking 
into account the errors in the total number of galaxies.   If we 
consider galaxies down to 10$^{5}$ this limit
becomes 38.9$^{+0.5}_{-0.7}$.

 Based on this calculation we predict that at around magnitude
35 we would find that the observed number counts of galaxies begins to
gradually decline
at fainter magnitudes, unless than are a significant number of galaxies with masses M$_{*} < 10^{6} M_{\odot}$ at the highest redshifts.  This depth however will be very difficult to reach
with even very deep surveys with JWST, and may require the next generation of
large space telescopes to fully probe.  This would also make deep observations
of future telescope observations confusion limited if they can reach these faint magnitudes.

If we want to observe all the galaxies in the universe, even with
imaging, it will require extensive telescope programmes that are likely
several generations away.  The Euclid mission planned for launch in 2020
will image 15,000 deg$^{2}$ down to AB magnitude of 24.5 in the wide VIS
filter which mimics in some ways the wide filter we construct in this paper
using the UDF-Max image (see Appendix).  The deep component of Euclid will 
be two magnitudes
fainter over 40 deg$^{2}$.  Using our number counts we predict that Euclid
will image around 3.7$\times 10^{9}$ galaxies in total, or $<$0.1\% of all
galaxies in the universe.  This is significantly more than we have imaged at
present with the Sloan Digital Sky Survey which has  imaged roughly 
$1.4 \times 10^{9}$ galaxies and QSOs, or $\sim 10^{-4}$ (0.05\%) 
of all galaxies 
in the universe.  LSST will find a similar number of galaxies.  
Thus we will have to wait at least several
decades before even the majority of galaxies have basic imaging in a single
band.  

\section{Summary}

We have investigated the fundamental question of the number density evolution
of galaxies in the universe.  We research this problem in a number of 
ways, and discuss
the implications for galaxy evolution and cosmology. We use recently measured 
 mass functions for galaxies up to $z \sim 8$ to determine the 
number density evolution of galaxies in the universe.
Our major finding is that the number densities of galaxies decrease 
with time such that the number density $\phi_{\rm T} (z) \sim t^{-1}$, 
where $t$ is the age of the universe.

We further discuss the implications for this increase in the galaxy
number density with look-back time for a host of astrophysical questions.
Integrating the number densities, $\phi_{\rm T}$ we calculation that there 
are $(2.0^{+0.7}_{-0.6}) \times 10^{12}$ galaxies in the universe up
to $z = 8$ which in principle could be observed.  This is roughly a factor
of ten more than is found through direct counting (see Appendix).
This implies that we have yet to detect a large population of faint 
distant galaxies.  

In terms of astrophysical evolution of galaxies, we show that the 
increase in the integrated
mass functions of all galaxies with redshift can be explained
by a merger model.   We show that a simple
merger model is able to reproduce
the decline in the number of galaxies with a merger time-scale of 
$\tau = 1.29\pm0.35$ Gyr.  The derived merger rate at $z = 1.5$
is $R \sim 0.05$ mergers Gyr$^{-1}$ Mpc$^{-3}$, close to the value found
through structural and pair analyses.   Most of these merging galaxies are 
lower mass systems based on the increase in number densities with time
seen at lower limit selections, using higher masses, for the total
number density calculation.  

We finally discuss the implications of our results for future surveys.  We
calculate that the number counts of galaxies at magnitudes fainter than
$m_{\rm max} = 29$ will largely probe the lower mass galaxies at higher 
redshifts and eventually at $m_{\rm max} \sim 35$ will turn over and 
decline due to 
reaching the limit of the number of galaxies in the universe, unless the
mass limit for galaxies is much less than M$_{*} = 10^{6}$ \solm or
there are many galaxies at $z > 12$.  We also show that this leads to a
natural confusion limit in detection and that these galaxies are likely
responsible for the optical and near-infrared background and provide a 
natural explanation for Olbers' paradox.  This large additional number of 
galaxies is also consistent with recent measures of
the cosmic infrared background light (e.g., Mitchell-Wynne et al. 2015).  

In the future, as mass functions become
better known with better SED modeling and deeper and wider data with JWST and
Euclid/LSST, 
we will be able to measure the total number densities of galaxies more precisely and thus obtain a better
measure of this fundamental quantity.

We thank Neil Brandt, Harry Teplitz, and Caitlin Casey for useful
discussions concerning non-optical deep detections in galaxy surveys.
This work was supported by grants from the Royal Astronomical Society,
STFC and the Leverhulme Trust.
Support was also provided by NASA/STScI grant HST-GO11082. A.M. 
acknowledges funding a European Research 
Council Consolidator Grant.

\appendix

\section{Direct Galaxy Counting in the Hubble Ultra-Deep Field}

In the Appendix we discuss how many galaxies can  be observed in the
universe directly with present-day instrumentation and telescopes using the Hubble 
Ultra Deep Field.   By doing this we address
how many galaxies can be detected with our present technology using
the deepest observations taken to date at all wavelengths from the
X-ray to the radio.   We do this to compare with the model results in the main
paper on the number density evolution of galaxies through cosmic time.
 Effectively our major results in the Appendix are the number of galaxies per
unit magnitude we can observe,  and quantifying how many galaxies we 
could detect if we observed the entire
sky at the same depth as the Hubble Ultra-Deep Field, ignoring the
effects of galactic extinction.   Although popular press releases have
discussed this number in the past, there has never been a published
version of this calculation.  

The Hubble Ultra Deep Field (HUDF) image was taken with the the 
Advanced Camera for Surveys (ACS) on the Hubble Space Telescope 
(Beckwith et al. 2006) (\S 3.1).    The HUDF images are taken in the
bands: F435W (B$_{435}$), F606W (V$_{606}$), F775W ($i_{775}$), and 
F850L ($z_{815}$).   The central wavelengths of the 
filters we use, and their full-width at half-maximum,
are: F435W (4297 \AA, 1038 \AA), F606W (5907 \AA, 2342 \AA), 
F775W (7764 \AA, 1528 \AA), F850L (9445 \AA, 1229 \AA).  

The original HUDF imaging program used 400 orbits of Hubble time for a
total exposure time of just under 1 Msec.  Later additions of data
were added over the years by programmes such as  Bouwens et al.
(2012).   The field of view of the ACS image for the HUDF is 11 arcmin$^{2}$. 
 Detailed examination of the HUDF imaging from previous work shows that 
the original
magnitude limit for point sources is m$_{\rm AB} \sim 28.5$ in the $z-$band at
8 $\sigma$ depth  using a 0.5\arcsec aperture (e.g., Beckwith et al. 2006), 
with increased data driving this depth even fainter.   This makes the
HUDF easily the deepest optical imaging taken to date.

The result for the number of galaxies that can be observed with
current technology which we calculate is largely based on these HUDF 
observations  and through using simulations to carry out corrections for
incompleteness.    Estimates
of this number from simply counting in deep fields has been provided in 
the past (e.g., Beckwith et al. 2006; Coe et al. 2006), yet we carry out 
this analysis in a different and more complete and careful way.   

To do this calculation, we create a maximal depth HUDF image from which
we carry out our counting analysis.  We  are also very
careful to not only detect the faintest galaxies, but also to ensure that
single galaxies are not over separated into separate systems or `shredded' by
the source extraction.


We supplement these counts with lower redshift counting where the 
HUDF area is not large enough to obtain a representative volume, 
as well as considerations for galaxies which are not detectable in observed
optical wavelengths.   First, we discuss in this section the number of galaxies
retrieved through direct detections in a combined maximum depth optical 
HUDF image, where the bulk of galaxies in this field, which can be detected, 
are found.  

\begin{figure}
 \vbox to 150mm{
\hspace{3cm} \includegraphics[angle=0, width=90mm]{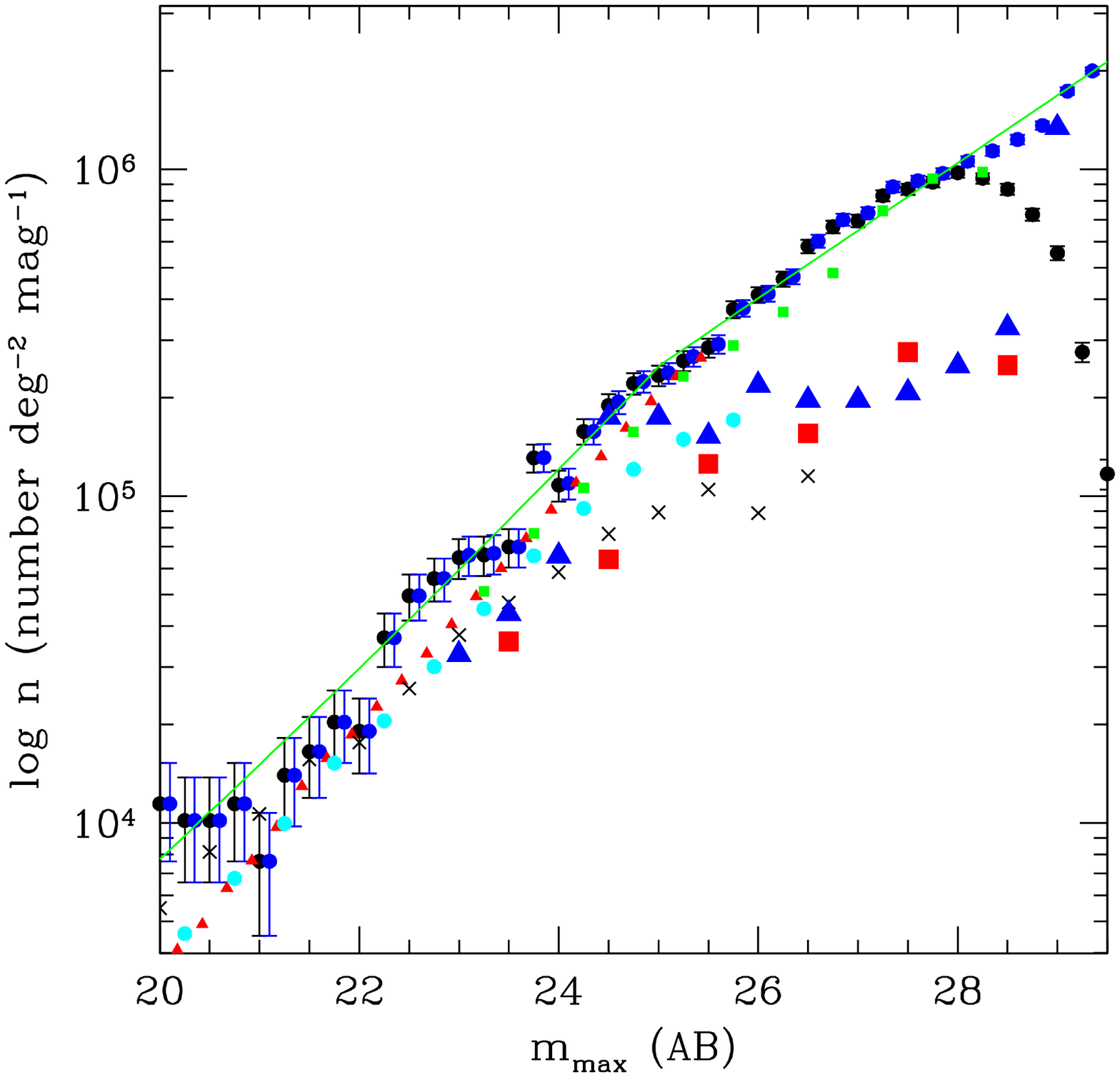}
 \caption{Plot showing the number counts of galaxies (n) within the
UDF-Max as a function of magnitude (see \S A.1.2).  The black points show
the raw counts of observed galaxies as a function of magnitude
while the blue circle points show the corrected counts for exponential like
profiles. Also shown here for comparison are number counts measured in
the red and infrared.  The small red triangles are from Capak et al. (2007) using
the i-band from the COSMOS survey,
while the cyan circles are from Metcalfe et al. (2006).  The black crosses are
from the H-band GOODS NICMOS Survey (GNS) (Conselice et al. 2011), the
red boxes are from the Hubble Deep Field South H-band (Metcalfe et al.
2006) and the large blue triangles are from the Hubble Deep Field
North (Thompson et al. 1999).  Note how our points from the UDF-Max 
are all shifted to brighter
fluxes, meaning that for the same galaxy, the magnitude is brighter in the
HDF-Max than in any individual band.}
} \label{sample-figure}
\end{figure}

\subsection{The Maximum Ultra Deep Field Image  (UDF-Max)}

To maximize the signal to noise of the HUDF imaging we combine the four
imaging bands - $BViz$ into one band which we call the 
``UDF-Max'' image.   We add
these images together according to their relative weights of signal to noise
(S/N) from their individual weight maps, and after ensuring that the PSF is 
matched between the bands.   The relative weights are determined from the 
depth of observations in
each band, such that the combination produces the highest possible signal
to noise ratio in the resulting combined data.  That 
is, these are added together according to their depth so as to maximize the
signal to noise of the average galaxy.  This will affect galaxies which
are only detected in the redder/bluer bands, but we later discuss how we
can ensure that these very red galaxies are still detected.    The formula
which we use to create this maximum depth image is given by:

\begin{equation}
{\rm UDF-Max} = 1.95 \times {\rm B} + 0.924 \times {\rm V} + 1.98 \times i + 4.12 \times z
\end{equation}

\noindent Where UDF-Max is the new HDF-Max image, and BV$iz$ are the
filters which are combined together.  We then 
recalibrate this combined image to obtain a new zero point of 
M$_{0}$ = 26.4 in AB magnitude units.  This was done by converting counts 
per second in each band to a flux and then determining the zero point based 
on this flux combination.  We then compute magnitudes in our constructed
UDF-Max image as follows:

\begin{equation}
{\rm m}_{\rm max} = -2.5\,{\rm log}\, C_{\rm max} + 26.4
\end{equation}

\noindent where $C_{\rm max}$ is the counts per second on the UDF-Max
image.  These magnitudes are then measured, and later used to calculate the
total number of galaxies using the number counts of galaxies at each 
magnitude.  We furthermore checked that all the objects
detected in each individual band are also detected in the combined image.
This is to ensure that, for example, galaxies detected only in the $z$-band,
as e.g., high redshift drop out galaxies, are also detected in the 
UDF-Max image.  This was indeed the case for all galaxies down to the detection
limit in each band.  

After creating this optimal maximal deep image, we then detect the galaxies 
within it. Our method of finding galaxies in the UDF-Max, as well
as for determining our completeness corrections, are done using the
SExtractor detection package (Bertin \& Arnouts 1996) using AUTO-MAG
magnitudes.    A simple detection
run through on the z-band image finds that there are between 9,000-10,000
galaxies within the Hubble Ultra Deep Field (see also Beckwith et al. 2006), 
but this does not take into account incompleteness. 
To optimize the number of galaxies we detect, and are able to successfully
extract and deblend from each other, we adjust the SExtractor 
detection and separation
parameters accordingly.  We do this through an iterative process whereby
we change the SExtractor detection and deblending parameters such that 
galaxies are SExtracted independently from each other.  That is, we 
repeat our detection methods until  
the faintest galaxies are detected and deblended.  This is done while at
the same time ensuring that noise is not being detected
as false positive objects.

Ultimately we detect our objects in the UDF-Max using a threshold of 
1.7$\sigma$, and a minimum detection area of 5 pixels. The deblending 
parameters responsible for separating 
nearby objects from each other are set to 32 deblending sub-thresholds 
with a minimum contrast parameter of 0.05.  We also combine the weight 
images from each of the four filter observations as above to remove 
noise which is otherwise detected as real objects by SExtractor.
We find using our methods described above that there are 9,713 galaxies
detected in the UDF-Max.   The distribution of the counts of the galaxies
in the UDF-Max at a given magnitude are shown as black circles in 
Figure~A1. If we slightly tweak
this value due to changing the detection parameters we get a variation on the
few percent level in the number of galaxies detected.

However, this is just a raw number, and to obtain the total number of
galaxies which are observable down to the depth of the UDF-Max, 
we must correct this measurement for the incompleteness of our observations.  
As such we carry out a series of simulations, matching the properties of 
high redshift galaxies, to determine the fraction of systems we are missing
when carrying out these measurements.

\subsubsection{UDF-Max Completeness}

Figure~A2 shows the detection completeness from simulating galaxies within our
UDF-Max image.    We compute this by 
simulating galaxy images at a given angular size, magnitude, and 
surface brightness profile as fake images, and then randomly
placing these into the UDF-Max image.  We then 
detect these with the same SExtractor method and parameters as 
we use for the detection of the real galaxies.  The completeness is then
defined as the number of simulated galaxies detected divided by the
total number put into the image at a given magnitude. 

When carrying out these simulation, we ensure that the angular size 
distribution at a given magnitude of the objects we input into 
each simulation  is the same as the observed distribution from our 
original SExtractor detections.   To carry this out we measure the 
angular sizes of our galaxies through Kron radii measured
through SExtractor, which we then later use in our distribution for 
the simulated galaxy angular sizes. We do this by 
fitting the observed angular size-magnitude relation and then simulating 
these galaxies such that they follow 
the same relationship, and to have the same scatter as measured 
statistically.  

At the faintest magnitudes we assume that the size distribution is 
similar in shape to brighter magnitudes, but with a different average value as
suggested by the relationship between magnitude and size, 
so as to not bias against 
large but fainter systems which would not be detected in our imaging. 
That is, we assume that the slope of the relationship between galaxy 
magnitude and size remains in place to our faint limit.  Thus
we are allowing for sizes of systems that would not ordinarily be detected
at the highest redshifts due to surface brightness dimming by
assuming a similar distribution of sizes as seen at slightly lower redshifts.

We carry out these simulations using different mixtures of 
surface brightness profile types for the simulated  
galaxies.  We then place these simulated galaxies in the UDF-Max 
after convolving with the PSF.  We then measure
the completeness of these simulations when we use pure elliptical (i.e., 
Vaucouleur) profiles for all galaxies, pure exponential disk profiles, 
and a 50/50 split of both.  Observations show that most high-z galaxies
have flatter profile shapes (e.g., Ferguson et al. 2004; 
Ravindranath et al. 2004; Buitrago et al. 2013), although we explore this 
range as a source of error.   Each of these simulations
is carried out, and the detection and incompleteness is calculated. This spans
the possible range of intrinsic galaxy profiles.  Based on this, the
completeness of our imaging drops quickly at magnitudes fainter
than m$_{\rm max}$ = 28, and by
magnitude 30 very few galaxies are still retrieved from our detection
methods.  This is similar  in pattern 
to the individual bands of the HUDF as shown by Beckwith et al. (2006).

\begin{figure}
 \vbox to 120mm{
\hspace{3cm} \includegraphics[angle=0, width=90mm]{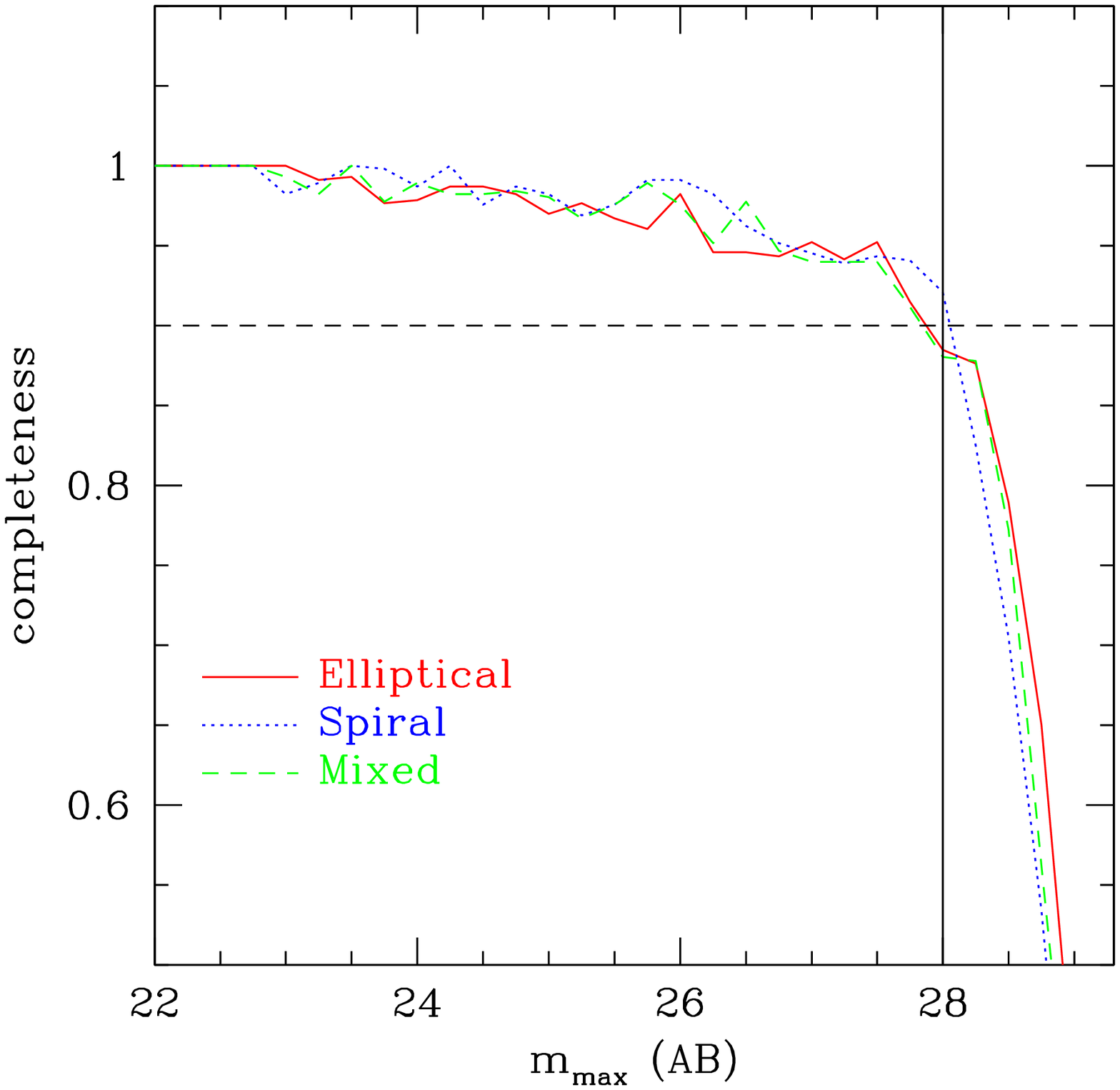}
 \caption{The detection fraction for galaxies within the
UDF-Max based on simulations of placing fake galaxies of the
same size distribution as objects within the UDF-Max and
using three different light profile shapes.  The red solid line
show systems which have light profiles of ellipticals, the
blue dotted shows disk profiles, and the green dashed line is an equal 
mixture of both.}
} \label{sample-figure}
\end{figure}

\subsubsection{Correct Observed Numbers}

To account for the incompleteness in our detected galaxies, and thus
to obtain a true total number of galaxies in the
UDF-Max, we divide the completeness into the number of observed
galaxies at a given magnitude.  The result of
the number counts before and after using this incompleteness correction
is shown in Figure~A1.    We also show the best fitting relationship between
the number density and magnitude as a green line.  This best fitting relation
has a break at $m_{\rm max}$ = 25.    The best fitting relation at 
$m_{\rm max} < 25$ is

\begin{equation}
{\rm log}\, n(m) = 0.048\pm0.001 \times {\rm mag}^{1.47\pm0.05}
\end{equation}

\noindent while for mag $> 25$ the relation is

\begin{equation}
{\rm log}\, n(m) = 0.244\pm0.003 \times {\rm mag}^{0.96\pm0.03}
\end{equation}

\noindent We then use these relations to calculate the total number 
density down 
to magnitude 29, where our detection completeness is $> 50$\%.  By 
integrating our magnitude fits in the UDF-Max from m$_{\rm max} = 20$ 
to 29, we calculate that the total number densities of galaxies, based
on just the UDF-Max image is $5.7^{+0.8}_{-0.7} \times 10^{6}$ 
 galaxies deg$^{-2}$ which is equivalent to $\sim 1600$ galaxies 
arcmin$^{-2}$ or $\sim 0.43$ galaxies arcsec$^{-2}$.
Using just this number density in the UDF-Max, we calculate the
total number of observable galaxies in the universe at optical 
wavelengths as:

$${\rm N_{tot,obs,opt}} = 2.34^{+0.36}_{-0.32} \times 10^{11} {\rm galaxies}$$ 

\noindent This uncertainty is solely based on number counting statistics, 
and as such does not account for other sources of error such as nearby 
and very distant ($z > 8$) galaxies,
non-optical detections and cosmic variance, which we discuss later
in the following sections.  Our number also does not differ substantially from
what can be derived from the original number counts from the HDF and UDF
(e.g., Williams et al. 1996; Beckwith et al. 2006).

These are the results when we simulate our galaxies as systems 
with purely disk-like profiles.  This is in fact a more accurate
description of distant galaxies, as disk-like profiles
become increasingly common at higher redshifts, including for the most
massive systems 
(e.g., Ferguson et al. 2004; Ravindranath et al. 2004; Buitrago et al. 2013).  
To obtain a measurement of  our uncertainties we recalculate
this number for simulated profiles which are 100\% elliptical like profiles, 
and for a 50/50 split between disk like systems and elliptical-like ones.  
We find using these different simulation corrections that the total number of 
galaxies is 2.36 and 2.57 $\times$ 10$^{11}$ respectively, with very
similar number counting errors as found in the pure disk-like simulations.
We therefore take the average of these two and use the difference
as another source of error. This gives us a total number of galaxies 
of 2.46$^{+0.38}_{-0.34} \times 10^{11}$ galaxies.

This is however an underestimate of the number of galaxies potentially
observable.  We have
also not yet taken into account the fact that the Hubble Ultra Deep Field does
not include the most distant objects in the universe which are
undetected in the optical, as well as the nearest galaxies which are
absent due to the nearby volume probed being very small in the UDF-Max. 
Although these numbers are small and make little difference to the overall
total number of galaxies, we include them for completeness.

\subsubsection{Ultra-high redshift and nearby galaxies}

We include the nearest galaxies in our calculation based on stellar mass
functions measured for local galaxies, as well as the deepest HST data
in the WFC3 to measure the ultra-high redshift systems which may exist
at $z > 8$.  

We use one of the latest deepest ultra-high redshift searches from the UDF12
project from Ellis et al. (2013).  This is a medium band survey of the
UDF done with the WFC3 near infrared camera.  By going very deep
WFC3 has made it easier to find ultra high redshift galaxies
than can be found within the ACS imaging. This  deep WFC3 data has
allowed Ellis et al. (2013) to identify in total seven galaxies between 
redshifts $z = 8-12$ which were previously 
unidentified.   Assuming this is a representative number density across
the sky, the total number of galaxies which are visible
to us at this redshift range is 2.2$\pm0.03 \times 10^{8}$ galaxies.

For galaxies in the nearby universe ($z \sim 0$) we use the mass 
functions calculated in the GAMA survey by Baldry et al. (2012).  
The effective mid-redshift for the
143 deg$^{2}$ covered in this study is $z \sim 0.03$ which
includes 5210 galaxies.  The derived stellar mass function is best
fit by a double Schechter function. Integrating this we calculate
the number density within the mass limits of M$_{*} = 10^{6} - 10^{12}$ \solm.
Doing this calculation we find
that there are $\sim 2.9\times 10^{8}$ galaxies at $z < 0.3$ that are 
not accounted for in the UDF-Max imaging.  Likewise, we also remove the
small number of galaxies in the UDF-Max imaging at $z < 0.3$ so as to 
not double count the number of galaxies at these lowest redshifts.

\subsubsection{Non-optical Wavelength Detections}

Whilst nearly all galaxies in the universe at $z < 6$ are detectable 
within observed optical light, as far as we know, it is possible or even 
likely that there
are galaxies at other wavelengths that cannot be detected in the
optical and near-infrared. 

First we investigate the number of galaxies which are detected at the sub-mm
and in the radio, but which are seen neither in the optical nor in the 
near-infrared.   The latest
surveys of sub-mm galaxies from Alberts et al. (2013) find that 85\% of
all sources have optical or near-infrared counterparts in deep GOODS 
imaging.  Using the number
densities from Casey et al. (2014), we find that between 0.1-1 mJy there
are 6.2$\pm0.05 \times 10^{8}$ extra galaxies across the sky
not matched to optical or near-infrared detections. Although the GOODS
data is almost as deep as the UDF, we cannot be sure that these sources
would not be found on the slightly deeper data, so we do not include them
in our total number.

In terms of ultraviolet detections, a recent survey by Teplitz et al. (2013) 
took very deep UV imaging of the Hubble Ultra Deep Field.  Matching to the deep
B-band HUDF data Teplitz et al. (2013) find that there are no systems in the
ultraviolet which do not have an optical counterpart.  Extragalactic
X-ray sources are similarly almost always detected in the optical/NIR.
The currently deepest X-ray data find that 716 out of 740  (96.8\%) of X-ray 
sources down to $\sim$ 3 $\times 10^{-17} {\rm erg cm^{-2} s^{-1}}$ over 
464.5 arcmin$^{2}$ at 0.5-8 keV are detected in deep optical data as 
part of the deepest 
4 Msec Chandra imaging of the universe (Xue et al. 2011) in the HUDF area.    
This gives us a
total of 7.67$\pm 1.6 \times 10^{6}$ X-ray sources in the universe that
are not accounted for by an optical and/or near infrared counterparts at
the UDF-Max depth.  This is much smaller than the sub-mm or optical drop 
out high$-z$ galaxies, and thus are not
likely a major contributor to the number of galaxies in the universe.

Therefore, in total we find that there are an additional 
$\sim 10^{9}$
galaxies that must be accounted for that are not detected in the optical
at $m < 29$.  This is however only 1\% of the nominal value we
obtained for the optical detections.   This implies that around 1\% of
galaxies that can be seen over all wavelengths are not detectable within 
deep Hubble Space Telescope imaging to the
depth of the Hubble Ultra Deep Field.

Our final number of galaxies
currently detectable within the universe with a hypothetical all sky
HST survey at UDF-Max depth is therefore 

$${\rm N_{tot,obs,cor}} = 2.47^{+0.38}_{-0.34} \times 10^{11}.$$ 

\noindent This results however does not accounting for cosmic variance 
which we discuss in the next section, and which is an added source
of uncertainty on this measurement.

\subsection{Cosmic Variance}

One of the things we must consider in an analysis such as this is the fact
that we are only probing a very small part of the sky when examining the 
UDF-Max.   It is certain that we would obtain a different answer if we were
to repeat this experiment by investigating galaxy counts in another
unrelated part of the sky/universe, obtaining a slightly higher or smaller 
number of counts.   This is due to the fact that the distribution of galaxies
can vary significantly as structure is highly clustered, and thus some
regions are more over dense than others.   

We can determine the contribution of this unknown random error to our results
by considering both the variance due to Poisson counting errors, as well as
this cosmic variance by using halo models of the universe.  To do this we 
utilize the cosmic variance code written by Trenti \& Stiavelli (2008) to 
determine for our observed sample the combined variance due to Poisson 
uncertainty and to cosmic variance.

The total variance changes significantly with the galaxy type and redshift.
At the highest redshifts, this relative variance can be as much as 0.31, 
while at lower redshifts it is around 0.05. The variance tends to decline at 
lower
redshifts as more representative galaxies are being studied, which lowers the
amount of cosmic variance.  Taking this into account in total we calculate
that the cosmic variance on our measured counts is 1.53$\times 10^{10}$ 
galaxies.   This then gives a combined uncertainty of 

$$N_{\rm tot,obs,cor \pm CV} = 2.47^{+0.41}_{-0.37} \times 10^{11},$$

\noindent which is our final estimate of the number of galaxies we can
observe today with current technology at HUDF depths in 
the  universe (ignoring magnification 
by lensing and extinction) with uncertainties
included for non-optical detections and cosmic variance.



\listofchanges

\end{document}